\documentclass[preprint,3p]{elsarticle}
\usepackage{graphicx}
\usepackage{xcolor}
\usepackage{multirow}
\usepackage{hyperref}
\usepackage{svg}
\usepackage{array}

\newcommand{\ssa}{\ensuremath{{}^{\textcolor{black}{\circ}}}} 

\makeatletter
\def\ps@pprintTitle{%
  \let\@oddhead\@empty
  \let\@evenhead\@empty
  \let\@oddfoot\@empty
  \let\@evenfoot\@oddfoot
}
\makeatother

\begin{document}
\begin{frontmatter}
\title{Utilizing ChatGPT to Enhance Clinical Trial Enrollment}

\author{Georgios Peikos\corref{cor1}\fnref{label1}}\ead{georgios.peikos@unimib.it}
\author{Symeon Symeonidis\fnref{label2}}\ead{symesyme@pme.duth.gr}
\author{Pranav Kasela\fnref{label1}}\ead{pranav.kasela@unimib.it}
\author{Gabriella Pasi\fnref{label1}}\ead{gabriella.pasi@unimib.it}

\cortext[cor1]{Corresponding author.}

\affiliation[label1]{organization={University of Milano-Bicocca, Department of Informatics, Systems and Communication},
            addressline={Viale Sarca 336}, 
            city={Milano},
            postcode={I-20126 MI}, 
            country={Italy}}
\affiliation[label2]{organization={Democritus University of Thrace, Department of Production and Management Engineering},
            addressline={Vasilissis Sofias 12}, 
            city={Xanthi},
            postcode={67100}, 
            country={Greece}}

\begin{abstract}
Clinical trials are a critical component of evaluating the effectiveness of new medical interventions and driving advancements in medical research. Therefore, timely enrollment of patients is crucial to prevent delays or premature termination of trials. In this context, Electronic Health Records (EHRs) have emerged as a valuable tool for identifying and enrolling eligible participants. In this study, we propose an automated approach that leverages ChatGPT, a large language model, to extract patient-related information from unstructured clinical notes and generate search queries for retrieving potentially eligible clinical trials. Our empirical evaluation, conducted on two benchmark retrieval collections, shows improved retrieval performance compared to existing approaches when several general-purposed and task-specific prompts are used. Notably, ChatGPT-generated queries also outperform human-generated queries in terms of retrieval performance.
These findings highlight the potential use of ChatGPT to enhance clinical trial enrollment while ensuring the quality of medical service and minimizing direct risks to patients.
\end{abstract}

\end{frontmatter}
\section{Introduction}
\label{sec:intro}
Clinical trials are the established scientific approach for assessing the effectiveness of new biological agents, drugs, devices, or procedures in preventing or treating diseases in human populations~\cite{fleming1996surrogate}.
Recruiting a sufficient number of patients to participate in a clinical trial is one of the main encountered challenges, as it not only causes delays and leads to trials' failure, but also compromises the validity of the conducted studies by limiting their generalizability~\cite{gul2010clinical,penberthy2012effort}.
The process of enrolling participants in clinical trials is intricate and comprises various steps, as reported in~\cite{jain2019conceptual}.
Typically, the process is initiated when a healthcare provider actively searches for an appropriate clinical trial for a certain patient.
The healthcare provider searches by utilizing the patient's clinical or genomic data derived from an Electronic Health Records (EHRs), which encompasses laboratory reports, radiology reports, or clinical notes~\cite{DBLP:journals/kais/LandolsiHR23}.
The search yields a list of clinical trials where the studied patient may meet their eligibility requirements.
At this point, human effort is necessary to refine the potential trial list and generate meaningful trial recommendations for the considered patient.
Once a suitable trial is identified, the results are shared with the patient's provider, who decides whether to proceed with this trial and requests a detailed patient prescreening to be performed.
After completing the prescreening process, the patient is contacted and offered the option to enrol in the trial. 
Patients who accept undergo a final screening to evaluate their eligibility, followed by a consenting process.
If the screening is successful and the patient consents, they are officially enrolled in the clinical trial.

EHRs have emerged as the preferred and effective approach for identifying and enrolling participants in clinical trials, complemented by strategies like reaching out to past participants and reviewing upcoming clinic schedules~\cite{hersh2007adding,o2021use}.
Empirical evidence indicates that incorporating EHR-based patient-screening in this task's workflow enhances recruitment rates~\cite{effoe2016use}.
However, according to~\cite{o2021use}, the lack of research-focused EHR-based modules restricts the optimal utilization of EHRs in recruitment efforts.
Two systematic reviews~\cite{von2021application,chow2023use} aim to comprehensively explore end-to-end applications of artificial intelligence (AI) in clinical trial enrollment by analyzing various research studies.
Their findings highlight the widespread utilization of Natural Language Processing (NLP) for extracting information from unstructured EHRs, the significant time savings achieved compared to manual screening methods, and the critical considerations of maintaining patient confidentiality and data security.
Another study further emphasizes the importance of leveraging NLP in clinical trial recruitment while emphasizing the need to assess their real-world adoption and effectiveness~\cite{DBLP:journals/jamia/IdnayDWS21}.
These insights further reinforce the notion that EHRs, with their abundance of patient-related information, including clinical narratives such as clinical notes, hold great potential for facilitating clinical trial enrollment.
However, as their primary purpose is to support clinical care rather than clinical trial enrollment, these narratives can be lengthy, unstructured, or contain several textual peculiarities, such as medical jargon and abbreviations.
To handle this textual limitations, several approaches in the literature exploit rule-based, hybrid, or neural-based NLP methods to extract valuable information from clinical narratives as highlighted in~\cite{DBLP:journals/kais/LandolsiHR23}.
In Section~\ref{sec:related_work}, we provide a comprehensive review of relevant research studies.

Our work focuses on the initial step of the workflow presented above, where healthcare providers actively search for eligible clinical trials by leveraging a patient's EHR.
We propose the usage of a large language model, namely the GPT-3.5 model, commonly referred to as ChatGPT\footnote{\href{https://openai.com/blog/chatgpt} {Introducing ChatGPT, accessed on 12/4/2023.}}, to extract patient-related information from clinical narratives.
In our approach, a clinical note that contains various patient-related information (e.g. underline medical problem, family history, patient's demographics) is processed through ChatGPT aiming to automatically synthesize queries for searching eligible clinical trials for the considered patient.
Therefore, our approach has the potential to fully automate this workflow step or, at least, significantly reduce the required human effort.

The extraordinary generative capabilities exhibited by various LLMs have lead to extensive discussions regarding the exploration and adoption in the medical domain.
One promising area of application for LLMs, such as ChatGPT, is the generation of discharge summaries~\cite{patel2023chatgpt}.
Due to their standardized format, discharge summaries could benefit from the utilization of ChatGPT, which has the potential to enhance the quality of these summaries.
An empirical investigation on the usage of ChatGPT has been conducted by medical experts~\cite{DBLP:journals/jms/CascellaMBB23}.
Their findings support the idea that ChatGPT can be utilized for generating medical notes, given an adequate amount of patient-specific information.
Furthermore, they outline that ChatGPT has the potential to effectively handle complex data and extract valuable information from various medical texts, including EHRs, clinical notes, and research papers.
However, the primary constraint of ChatGPT lies in its capacity to address causal relationships among conditions and comprehend the intricate connections between different conditions and treatments.
Another possible application of an LLM model, specifically the GPT-3 model, in a medical context is presented in~\cite{sezgin2022operationalizing}.
The authors discuss the implementation and operationalization of GPT-3 in clinical practice, focusing on factors such as integration with existing hospital networks, ensuring secure connectivity, incorporating text summarization services, and storing generated information in patients' EHRs.
Although LLMs demonstrate promising capabilities, there are limitations that give rise to concerns regarding their adoption in the medical domain~\cite{harrer2023attention,albrecht2022despite,patel2023chatgpt,DBLP:journals/jms/CascellaMBB23}.
Besides their well-known limitations such as, lack of understanding and reasoning, hallucination generation~\cite{ji2023survey}, inconsistent responses, outdated knowledge, their application in the medical domain also raises reliability and safety concerns~\cite{albrecht2022despite}.
For instance, the ability of ChatGPT to generate answers that may sound plausible but are incorrect raises concerns about the possibility of  providing incorrect information regarding medication changes, which can significantly impact patient care~\cite{patel2023chatgpt}.
Having said that, in our study, ChatGPT is employed as an intermediate step in a retrieval pipeline, while its role is to extract patient-related information from unstructured clinical notes rather than generate new pieces of information or make any decisions.
Hence, our approach has a reduced exposure to the aforementioned limitations typically associated with LLMs.

In this study, we propose an approach that automates the initial step of the enrollment process while maintaining the quality of medical service and minimizing direct risks to patients.
We leverage ChatGPT to extract patient-related information from unstructured clinical notes and generate search queries for retrieving potentially eligible clinical trials.
This study aims to address the following research questions:
\begin{enumerate}
    \item Is the patient-related information extracted by ChatGPT sufficient to improve retrieval performance?
    \item Which of the employed prompting approach yields the highest retrieval performance in the studied search task?
    \item Does the utilization of ChatGPT enhance retrieval performance compared to existing state-of-the-art approaches in the literature?
    \item What is the achieved retrieval performance of queries generated by ChatGPT and those generated by humans?
    \item Can the integration of ChatGPT into the clinical trial enrollment pipeline prove beneficial?
\end{enumerate}
This study has been organised in the following way.
In Section~\ref{sec:lit_rev}, a thorough review of relevant studies is presented, encompassing key areas such as prompting and in-context learning techniques, medical information extraction approach, LLMs used for query generation, and clinical trial retrieval methodologies introduced in the literature. 
In Section~\ref{sec:our_approach}, we present our proposed application of ChatGPT as an intermediate step in a retrieval pipeline focused on the task of clinical trials retrieval.
We explore its capabilities in three key areas: query generation, information extraction, and entity meaning disambiguation. Furthermore, we discuss how we integrate ChatGPT with a transformer-based model to handle the task of negation removal.
This section provides a clear explanation of how we leverage ChatGPT to improve medical information extraction.
Section~\ref{sec:exp_setup} focuses on the conducted experiments, thoroughly examining ChatGPT's performance.
This section provides valuable insights into the experimental setup and sheds light on the capabilities and effectiveness of ChatGPT in handling the required information extraction tasks.
The overall findings and a complete discussion are presented in Section~\ref{sec:find_disc}, where we compare our approaches with state-of-the-art methods and human-generated queries. 
Section~\ref{sec:limitations} addresses the limitations and potentials of using ChatGPT to enhance clinical trials retrieval.
Finally, in Section~\ref{sec:conclusions}, we conclude the paper and present directions for future research.
\section{Literature Review}
\label{sec:lit_rev}
This section reviews research conducted on LLMs, specifically emphasizing in-context learning settings and prompting techniques. First, we provide an extensive overview of advancements in the field of Medical Information Extraction, highlighting the transition from traditional rule-based approaches to the adoption of LLMs. Also, we discuss relevant studies that utilize LLMs as intermediary steps for query generation, aiming to improve retrieval performance across different search tasks. Finally, we narrow our focus to clinical trials retrieval, our research's primary focus.

\subsection{Language Models: From Prompting to In-context learning}
\textit{Prompting} is a technique that leverages language models to perform prediction tasks by directly predicting the probability of text.
In these models, the original input $x$ is modified using a \textit{template} into a textual string, i.e. \textit{prompt} $x'$, that contains some unfilled slots (often related to the desired prediction).
Then, the language model is used to fill the unfilled information and output a final string $x''$, from which the final output $y$ (desired prediction) can be derived~\cite{DBLP:journals/csur/LiuYFJHN23}.
Based on the literature, there are two distinct types of prompts, namely \textit{soft prompts} and \textit{hard prompts}.
Soft prompts are learned embeddings or vectors that guide the model towards generating desired responses~\cite{DBLP:conf/emnlp/LesterAC21}.
Hard prompts refer to human-engineered textual inputs that provide context, instructions, or examples to guide the model toward generating an appropriate response. 
In this work, the term \textit{prompts} refers always to \textit{hard prompts}, unless specified otherwise.

The development of LLMs like GPT-2~\cite{radford2019language}, GPT-3~\cite{DBLP:conf/nips/BrownMRSKDNSSAA20}, among others, allowed the use of prompts (i.e. hard prompts) that contain task-related instructions or demonstrations (i.e. task-specific input-output example pairs). These prompts are provided as natural language in the LLMs during inference time, and the models are expected to complete further instances of the task.
This process is referred to as ``in-context learning''~\cite{DBLP:conf/nips/BrownMRSKDNSSAA20}. 
In-context learning mainly leverages three techniques; ``few-shot learning'', where the prompt contains a few demonstrative examples (often between 10 to 100), ``zero-shot learning'' where only a task description with no tasks related examples is provided, or ``one-shot learning'' where one single example along with the task description is given to the model~\cite{DBLP:conf/nips/BrownMRSKDNSSAA20}.
 
The selection of the appropriate prompt for in-context learning is essential to the overall LLM's effectiveness in the considered task~\cite{zhao2023survey,DBLP:conf/nips/PerezKC21,DBLP:conf/chi/ReynoldsM21}.
Due to that, various research works introduce new prompting strategies for more efficient hard prompt construction, such as chain-of-though~\cite{DBLP:conf/nips/Wei0SBIXCLZ22},
least-to-most prompting~\cite{DBLP:journals/corr/abs-2205-10625},
instruction prompt tuning~\cite{DBLP:journals/corr/abs-2212-13138},
self-consistency~\cite{DBLP:journals/corr/abs-2203-11171},
and chaining multiple LLM prompts together~\cite{DBLP:conf/chi/WuTC22}.

Other works have been focused on prompt tuning, i.e. soft prompt construction or propose hybrid prompting approaches~\cite{DBLP:conf/emnlp/LesterAC21,DBLP:journals/corr/abs-2112-00114,DBLP:journals/corr/abs-1909-05858}.
Other research endeavors focus on addressing multiple limitations of LLMs, such as enhancing their capacity for reasoning~\cite{DBLP:journals/corr/abs-2205-10625,DBLP:journals/corr/abs-2205-09712,DBLP:conf/nips/KojimaGRMI22}.

Wei et al.~\cite{DBLP:conf/nips/Wei0SBIXCLZ22} introduced the Chain-of-Thought (CoT) method as a way to address the limitations of large language models in arithmetic, commonsense, and symbolic reasoning~\cite{DBLP:journals/corr/abs-2112-11446}.
The CoT prompting technique is a form of few-shot prompting that includes an \textless input, chain-of-thought, output\textgreater ~triplet.
The chain-of-thought component comprises a series of natural language reasoning steps, for instance, human-like thoughts of solving a mathematical problem, that guide the model to produce the desired output.
By applying this approach to three different LLMs, the authors showed its effectiveness in enhancing performance across arithmetic, commonsense, and symbolic reasoning tasks.
According to Zhou et al.~\cite{DBLP:journals/corr/abs-2205-10625}, chain-of-thought prompting performs poorly when the requested problem is more complex than those included as demonstrations in the prompt.
To overcome this, they proposed Least-to-Most prompting, a two-step approach that simplifies complex problems into more manageable sub-problems.
The first step, problem reduction, supplies the model with examples and a specific question to break them into sub-problems.
In the second step, problem-solving, the model sequentially addresses these sub-problems using constant examples, previously answered sub-questions, and generated solutions for guidance.
Occasionally, the two stages can be combined into a single-pass prompting. 
Experimental findings in symbolic manipulation, compositional generalization, and mathematical reasoning demonstrate that least-to-most prompting substantially surpasses both standard prompting and chain-of-thought prompting in performance.
Self-consistency~\cite{DBLP:journals/corr/abs-2203-11171} is another strategy that aims to improve the performance of chain-of-thought prompting.
The intuition behind this approach is that a complex reasoning problem often has multiple ways of thinking that lead to the same correct answer.
Therefore, by considering diverse reasoning paths and focusing on the most consistent answer, the self-consistency method aims to enhance the model's ability to solve complex reasoning tasks.
In this approach, the final answer is the one with the majority vote.
The empirical evaluation conducted by the authors suggests that the self-consistency method significantly improves the performance over the chain-of-thought prompting on a range of popular arithmetic and commonsense reasoning benchmarks.
Moreover, the method also outperforms the Least-to-Most Prompting approach on the arithmetic reasoning task, based on the results obtained using the GSM8K dataset~\cite{DBLP:journals/corr/abs-2110-14168}.
Wu et al.~\cite{DBLP:conf/chi/WuTC22} introduced the concept of Chaining LLM steps together.
In this prompting approach the output of one prompt becomes the input of the next, thereby combining the benefits gained at each step.
Through the utilization of Chaining, a complex problem is decomposed into various smaller sub-tasks, each associated with a separate prompt.
A user study conducted by the authors showed that the implementation of Chaining not only resulted in improved task outcomes but also contributed to users' satisfaction, sense of control, collaboration, and enhanced transparency of the LLM-based system.

Lester et al.~\cite{DBLP:conf/emnlp/LesterAC21} presented prompt tuning as a simple and computationally efficient technique for adapting LLMs to particular downstream tasks. This method involves learning soft prompt vectors through back-propagation while keeping the rest of the LLM frozen. 
During the tuning process, a task-specific copy of the entire pre-trained model is created for each downstream task, and inference is conducted in separate batches. As a result, prompt tuning requires only a small task-specific prompt per task, enabling mixed-task inference using the original pre-trained model.
The authors' experimental results indicate that adapting frozen pre-trained language models to downstream tasks using prompt tuning helps prevent overfitting to a specific domain.
Following the research work mentioned above, in~\cite{DBLP:journals/corr/abs-2112-00114}, the authors introduce instruction prompt tuning. This method combines a soft prompt learned through prompt tuning with a task-specific human-engineered hard prompt.
The authors evaluated the performance of their method in MultiMedQA multiple-choice datasets, and their approach surpasses prior
state-of-the-art by 17\%.
Another model that allows for task-specific adaptation is the CTRL model introduced in~\cite{DBLP:journals/corr/abs-1909-05858}.
It is trained with \textit{control codes} that can be related to a domain, subdomain, entities, relationships between entities, dates, and task-specific behaviour (e.g. question answering or translation).
As a result, the text generation process during inference is easily controlled by its end users. 

LLMs have already reached state-of-the-art (SoA) performance in various tasks, and selecting an appropriate prompt has a significant impact.
As a result, a significant number of related works investigate techniques to improve prompt construction further and their effectiveness~\cite{DBLP:conf/acl-deelio/LiuSZDCC22,rubin-etal-2022-learning,shin-etal-2021-constrained}, explore their robustness to permutations of the demonstrative examples~\cite{DBLP:conf/icml/ZhaoWFK021}, their sensitivity to negations~\cite{DBLP:journals/corr/abs-2209-12711}, and their ability to generalize across different LLMs~\cite{DBLP:journals/corr/abs-2302-09865}.

To begin with, Liu~\cite{DBLP:conf/acl-deelio/LiuSZDCC22} investigates the sensitivity of GPT-3's performance to the selection of in-context demonstrative examples.
The authors propose KATE, a retrieval-based approach for prompt construction that, given a test query (i.e. required question to the model), selects semantically-similar examples to the query and uses them to construct the final prompt.
Their findings suggest that this approach consistently outperforms random prompt selection on various NLP benchmarks, with notable gains observed in tasks such as table-to-text generation.
Similarly, Rubin et al.~\cite{rubin-etal-2022-learning} found that retrieving semantically-similar examples to the query and adding them in the final prompt improves effectiveness on three sequence-to-sequence tasks that map utterances to meaning representations.
Similar conclusions have been drawn in~\cite{shin-etal-2021-constrained}.
This evidence suggests that using semantically-similar demonstrative examples to the final query, rather than randomly selected, is a better practice for prompt construction.

Zhao et al.~\cite{DBLP:conf/icml/ZhaoWFK021} showed that GPT-3's few-shot learning could be unstable due to the selected prompt format, number of training examples, and example order. 
To address this issue, the authors introduced a contextual calibration procedure that significantly improves GPT-3 and GPT-2's accuracy and stability across various prompt choices.
LLM's sensitivity to ordering a prompt's examples has also been investigated by Lu et al.~\cite{lu-etal-2022-fantastically}, where the authors showed that it significantly impacts effectiveness across various NLP tasks.
Jang et al.~\cite{DBLP:journals/corr/abs-2209-12711} explored LLMs ability to comprehend and respond in negated prompts.
The authors experimented with several LMs and LLMs, e.g. GPT-3, InstructGPT~\cite{DBLP:conf/nips/Ouyang0JAWMZASR22}, among others and their findings suggest that scaling LMs does not enhance their ability to understand negations.
In certain situations, LLMs benefit from in-context learning to understand negation, while fine-tuning is effective in all scenarios. However, fine-tuning negatively impacts the performance of the original task. Despite these approaches, the LLMs used in the study still fell short compared to human performance.
Another study explored the behaviour of ChatGPT in terms of semantic, negation, and symmetric consistency~\cite{DBLP:journals/corr/abs-2303-06273}.
Semantic consistency implies that a model should make coherent decisions in contexts that have the same meaning.
Symmetric consistency is a type of consistency that relies on symmetric inference, meaning that for a given function $f$, if $f(x, y)$ yields a result, then $f(y, x)$ should produce the same result. Finally, negation consistency revolves around the logical negation property.
Their findings suggest that, ChatGPT exhibits improved language understanding, particularly in negation expressions and antonyms, compared to other LMs.
However, it displays self-contradictory behavior by frequently changing its predictions when presented with paraphrased inputs.
Finally, ChatGPT tends to generate different outcomes when the order of input sentences is altered, violating symmetric consistency.
The authors emphasize the significance of human inspection in AI-generated content, particularly for risk-sensitive applications, as revealed by their findings.

Rakotonirina et al.~\cite{DBLP:journals/corr/abs-2302-09865} investigated whether prompts can generalize across different LMs and LLMs focusing on the slot filling NLP task.
The authors experimented with manual, semi-manual, and automatic methods for prompt creation.
Their empirical evaluation suggests that prompts are generally more stable across different sizes of the same model.
They modified the AutoPrompt algorithm~\cite{shin2020autoprompt} so that one LM generates candidate prompts, and then a second model evaluates them and chooses the best one.
\subsection{Medical Information Extraction: From rule-based approaches to LLMs}
\label{sec:related_work}
Medical information extraction is a sub-field of natural language processing that focuses on extracting essential information from unstructured clinical text, such as electronic health records, clinical notes, medical literature, and patient narratives.
The primary goal is to improve medical decision-making, disease surveillance, clinical research, and personalized patient care.
To that aim, research interest has been focused on NLP tasks such as medical named entity recognition, relation extraction, event extraction, temporal information extraction, and negation detection, among others~\cite{DBLP:journals/kais/LandolsiHR23}.
The field of medical information extraction (IE) has experienced substantial development over the years, transitioning from rule-based approaches to supervised machine learning methods, progressing even further with the adoption of deep neural networks, transformer-based models, large language models, and, ultimately, domain-specific LLMs.
Early IE systems relied on manually crafted rules to identify and extract relevant clinical information.
Despite their limitations in scalability, for many tasks, especially involving extraction of numbers, acceptable performance was often achieved with relatively simple rule-based approaches~\cite{DBLP:journals/jbi/KreimeyerFPAHJF17,DBLP:journals/jbi/WangWRMSALZMSL18}.
The emergence of machine learning techniques along with the availability of domain-specific data sets accessible to the research community with a data-use agreement (e.g., i2b2\footnote{\href{https://www.i2b2.org/NLP/DataSets/} {National NLP Clinical Challenges, accessed on 12/4/2023.}} and MIMIC II~\cite{saeed2011multiparameter}), 
allowed more accurate and robust extraction of medical entities and their relationships~\cite{DBLP:journals/jamia/JiangCLRMDX11}.
With the advent of deep learning, models based on word embeddings that leverage recurrent neural networks, such as Recurrent neural networks (RNNs), enabled more effective representation and understanding of complex medical language~\cite{DBLP:journals/jamia/WuRDDJSSWWXZX20}.
Nonetheless, the limited availability of datasets has restricted the presence of deep learning approaches in non-English languages, such as French, resulting in a comparatively smaller adoption~\cite{DBLP:journals/jamia/WuRDDJSSWWXZX20}.
Transformer architectures, such as BERT~\cite{DBLP:conf/naacl/DevlinCLT19} achieved state-of-the-art performance on many generic NLP tasks, and following it, many clinical and biomedical variations, like ClinicalBERT~\cite{alsentzer2019publicly}, SciBERT~\cite{beltagy2019scibert}, among others, have been proven effective in domain-specific NLP tasks~\cite{landolsi2023information,hahn2020medical}.

The introduction of large language models like GPT-3, PaLM~\cite{DBLP:journals/corr/abs-2204-02311}, and GPT-4~\cite{DBLP:journals/corr/abs-2303-08774}, among others, has revolutionized the field of natural language processing~\cite{fan2023bibliometric}. Their pre-trained knowledge and fine-tuning capabilities have facilitated substantial progress in various NLP tasks, such as information extraction, summarization, and question-answering.
In~\cite{zhao2023survey}, the authors provide a detailed overview of four aspects of LLMs, namely the pre-training process (i.e. data collection, architectural design, and model training), adaptation tuning (i.e. effectively tune pre-trained LLMs), utilization (i.e. usage of LLMs to solve downstream tasks), and evaluation.
They highlight the main issues of LLMs, such as the problem of hallucination generation~\cite{DBLP:journals/corr/abs-2302-04023}, the inability to address tasks that require knowledge beyond the training data (i.e. knowledge recency), and the inconsistency in the provided answers, among others.
Finally, they discuss the potential risks and capabilities of LLMs that may arise within the medical domain.

In the medical domain, general-purposed LLMs have been employed to analyze electronic health records and clinical notes (i.e. unstructured clinical text) to aid the diagnostic process and offer treatment suggestions, among others~\cite{fan2023bibliometric}.
In addition, Liu et al.~\cite{DBLP:journals/corr/abs-2303-11032} investigated the usage of ChatGPT\footnote{\href{https://openai.com/blog/chatgpt}{Introducing ChatGPT, accessed on 12/4/2023.}} and GPT-4 in another task, i.e. medical text anonymization.
Their empirical evaluation showed that both of these models (in a zero-shot setting) are capable of 
de-identifying medical data compared to ClinicalBERT.
Regarding the employed prompts, they found that explicit prompt design that contains a well-written description of the desired output,
clearly defines the task, and provides concrete examples, leads to better performance.
As highlighted by Zhao et al.~\cite{zhao2023survey}, answer inconsistency is a significant issue of generative LLMs.
To solve this problem, Chuang et al. proposed SPeC, a model-agnostic soft prompt-based calibration pipeline that addresses the issue of output variance in clinical note summarization \cite{chuang2023spec}. By employing soft prompts along with discrete prompts, the proposed method effectively mitigates summarization variance while still harnessing the benefits of prompt-based summarization across three LLMs.
Since the introduction of LLMs and their adoption to solve specific NLP tasks in the health domain (biomedical and clinical), several studies have investigated whether these general-purposed LLMs are proper tools or if smaller, pre-trained models on domain-specific NLP tasks should be used instead~\cite{DBLP:journals/corr/abs-2302-08091,DBLP:journals/corr/abs-2205-12689,DBLP:conf/emnlp/GutierrezMWCLS022,DBLP:journals/corr/abs-2109-02555,hu2023zero,DBLP:journals/corr/abs-2303-04360}.
Now, we provide an overview of the aforementioned studies, by focusing on the investigated NLP task, the compared LLMs, and pre-trained/fine-tuned LMs.
In addition, we comment on the selected prompting, as it is highly related to the effectiveness of LLMs~\cite{zhao2023survey,DBLP:conf/nips/PerezKC21}. 

Lehman et al.~\cite{DBLP:journals/corr/abs-2302-08091} investigated whether LLMs can yield better effectiveness across three medical NLP tasks in clinical information extraction; two were related to the multi-label classification of clinical sentences, and one was related to medical question-answering.
To that aim, they compared 12 language models, i.e. BioClinRoBERTa~\cite{DBLP:conf/acl-clinicalnlp/LewisODS20}, GatorTron~\cite{DBLP:journals/corr/abs-2203-03540} (an LLM trained on de-identified clinical texts), PubMedGPT (which is now renamed to BioMedLM\footnote{\href{https://huggingface.co/stanford-crfm/BioMedLM}{BioMedLM, accessed on 12/4/2023.}}, GPT-3 and T5~\cite{raffel2020exploring}.
Regarding the prompts used with GPT-3, a single prompt was employed to simultaneously instruct the model to generate predictions for all labels.
Their findings suggest that models fine-tuned on all available data, particularly BioClinRoBERTa and GatorTron, significantly outperform any in-context learning approach for the selected NLP tasks.
Nonetheless, the authors did not employ ChatGPT in their evaluation as it is unavailable via a HIPAA-certified API.
Similar conclusions have been drawn from the empirical evaluation of Gutierrez et al.~\cite{DBLP:conf/emnlp/GutierrezMWCLS022} in biomedical information extraction.
In their work, the authors compare the few-shot performance of GPT-3 in-context learning with fine-tuning smaller LMs, namely PubMedBERT-base~\cite{DBLP:conf/emnlp/GutierrezMWCLS022}, BioBERT-large~\cite{Lee20} and RoBERTa-large~\cite{DBLP:journals/corr/abs-1907-11692}.
They investigate two biomedical NLP tasks, i.e., named entity recognition and relation extraction, across eight datasets, aiming to extract diseases, chemicals, medical concepts, and genes, identify drug-to-drug and chemical-to-protein interactions, and associate genes with diseases.
The authors paid particular attention to the in-context learning process of GPT-3 by following a systematic and task-agnostic process for constructing the prompts.
In detail, they constructed prompts based on the \textit{True Few-Shot} training process introduced by Perez et al.~\cite{DBLP:conf/nips/PerezKC21}, aiming to avoid plausible biases introduced in the model due to prompt selection on a large validation set.
Their evaluation suggested that GPT-3 significantly underperforms compared to the employed fine-tuned LMs.
Moradi et al.~\cite{DBLP:journals/corr/abs-2109-02555} investigated whether GPT-3 following a few-shot in-context learning setting outperforms BioBERT on various biomedical and clinical NLP tasks.
The prompts associated with GPT-3 contained a description of the task and a few examples, instructing the model on formulating its response. 
Their findings highlight GPT-3's inability to compete with BioBERT, especially in tasks that require calculating a semantic similarity score between sentences.
BioGPT~\cite{DBLP:journals/bib/LuoSXQZPL22} introduced by Luo et al., achieved the highest performance compared to GPT-2, and several other domain-specific LMs, when evaluated on six biomedical NLP tasks, namely, relation extraction, question answering, document classification and text generation. 
BioGPT is a domain-specific LLM with the same model architecture as GPT-2, and it is pre-trained on a 15M PubMed abstracts corpus.

The findings presented in the aforementioned studies~\cite{DBLP:journals/corr/abs-2109-02555,DBLP:conf/emnlp/GutierrezMWCLS022,DBLP:journals/corr/abs-2302-08091,DBLP:journals/bib/LuoSXQZPL22} suggest that BioGPT achieved state-of-the-art performance compared to pre-trained and fine-tuned LMs in biomedical NLP tasks, while GPT-3 based on in-context learning did not yield performance improvements.

Regarding clinical IE, the work of Hu et al.~\cite{hu2023zero} explored the potential of using ChatGPT for clinical named entity recognition in a zero-shot setting.
Their results showed that ChatGPT surpassed GPT-3 in terms of F1 scores for both exact- and relaxed-matching on an annotated subset of transcribed medical reports, i.e. MTSamples\footnote{\href{https://mtsamples.com/}{MTSamples, accessed on 14/4/2023.}}.
However, ChatGPT under performed compared to BioClinicalBERT\footnote{\href{https://huggingface.co/emilyalsentzer/Bio_ClinicalBERT}{Bio+Clinical BERT model, accessed on 14/4/2023.}} fine-tuned on the i2b2 2010 dataset~\cite{DBLP:journals/jamia/UzunerSSD11}.
The authors employed two types of prompts; for example, to extract medical problems, the first prompt was:
\textit{``Extract without rephrasing all medical problem entities from the following note in a list
format:''}; the second, which led to better performance, was:
\textit{``Extract without rephrasing all medical treatment, medical procedure, medical intervention, medication, drug entities from the following note in a list format:''}.
The performed error analysis revealed that ChatGPT might attempt to infer or summarize information or rephrase terms, even though it has been explicitly instructed not to.
Lastly, the authors note that ChatGPT's performance might have been underestimated due to minor changes in its response.
Agrawal et al.~\cite{DBLP:conf/emnlp/AgrawalHLKS22} investigated GPT-3's and InstructGPT's~\cite{DBLP:conf/nips/Ouyang0JAWMZASR22} ability to perform zero- and few-shot information extraction from clinical text.
To that aim, they compared their performance to various LM models. They showed that GPT-3 performs well in clinical NLP over diverse tasks, namely abbreviation expansion, coreference resolution, extraction of biomedical evidence, medication status, and medication attribute.
Based on the employed prompts, findings show that a \textit{guided prompt design} leads to performance improvements.
Similarly to Hu et al.~\cite{hu2023zero}, they found that GPT-3's outputs did not always match the annotated text (required output) at the token level, suggesting that its performance could have been greater.
In addition, they highlighted GPT-3's bias towards responding to a question, i.e. extracting a piece of information, even though the requested entity does not exist in the given text.
All in all, the previous studies showed that ChatGPT and GPT-3, two general-purposed LLMs, have the potential to perform accurate IE in the clinical domain (occasionally even outperforming domain-specific LMs) and highlighted some potential issues related to models' response behaviour (strong dependence on the created prompts) and their evaluation (mainly due to token-level miss-match).
Yang et al.~\cite{DBLP:journals/npjdm/0015CPSSPC0CFZM22} developed a large clinical language model from scratch, namely GatorTron. 
The model has adopted the BERT architecture with three different settings varying from the base model with 345M parameters to the large model with 8.9B parameters.
The model has been evaluated across five clinical NLP tasks: clinical concept extraction, medical relation extraction, semantic textual similarity, natural language inference, and medical question answering.
Empirical findings show that GatorTron outperforms previous transformer models, such as BioBERT and ClinicalBERT, across all NLP tasks.
However, as the authors mention, GatorTron achieved remarkable improvements for complex NLP tasks such as natural
language inference and medical question answering, but shows only marginal improvements in simpler tasks such as clinical concept extraction and medical relation extraction.

To conclude, LLMs have shown great potential for medical information extraction.
As the empirical evidence suggests, in the biomedical domain, general-purposed LLMs, like GPT-2, GPT-3, and ChatGPT, fail to reach the effectiveness of LMs in essential NLP tasks.
LLMs trained from scratch on domain-specific data, such as BioGPT, performed better than previous state-of-the-art approaches. 
In contrast, even general-purposed LLMs have improved performance for clinical information extraction over the previous SoA models like ClinicalBERT.
However, in both domains, the LLMs performance is highly related to prompt formulation and the models have been found to be very sensitive to that.
In addition, it has been reported that it might be the case that the performance of LLMs might have been underestimated due to their tendency to rephrase extracted tokens in their responses.
A limitation that has been identified that may play a crucial role is that in the majority of the works, the prompts followed a task-agnostic approach. 
Moreover, other model parameters, like ChatGPT's parameter related to the system's role, have yet to be fully investigated in the studies mentioned above.
Lastly, in the clinical domain, it has been found that LLMs like Gatortron and LMs like BioBERT perform similarly in simple entity extraction tasks.
\subsection{LLMs for Query Generation to Enhance Retrieval}
The intersection between Information Retrieval (IR) and Natural Language Processing (NLP) has become more prominent in recent years.
Some applications of LLMs in the field of IR involve the creation of synthetic datasets tailored to specific domains and tasks.
Specifically, a recent study has leveraged LLMs to generate synthetic training
datasets for IR tasks~\cite{DBLP:conf/sigir/BonifacioAFN22}. Their findings suggest that models fine-tuned exclusively on their synthetic datasets surpass strong benchmarks, including BM25, as well as recent self-supervised dense retrieval approaches.
Similarly, in~\cite{DBLP:journals/corr/abs-2303-00807}, the authors proposed UDAPDR, a strategy that uses synthetic queries created using generative models, such as GPT-3, to train multiple passage re-rankers on queries for target domain passages.
Their evaluation on three datasets showed that UDAPDR could improve zero-shot retrieval accuracy on new domains without using labelled training examples.

Another example of the synergy of NLP and IR is query generation or expansion, which is the main focus of this work.
Specifically, due to their vast accumulated knowledge, LLMs might be capable of paraphrasing or expanding queries and improving search quality, especially for standard retrieval models that rely on bag-of-words and are commonly used as first-stage retrievers.
In this setting, a query can be input into an LLM as a prompt, accompanied by task-specific instructions, allowing the model to generate contextually relevant and accurate responses (i.e. reformulated query) tailored to the information needed and the task to be performed.
This research direction has been investigated in~\cite{DBLP:conf/webi/Claveau21};
the proposed approach improves information retrieval using GPT-2 to generate multiple texts based on a given query.
The generated texts are concatenated to create an expanded query, providing broad coverage of vocabulary that captures synonyms, hypernyms, and other linguistic relations.
Then, the expanded query is used as an input in an IR system.
In the described process, the only online task is text generation, while model training and fine-tuning are performed offline.
The experiments conducted on several datasets showed the effectiveness of this approach over other query expansion methods, such as RM3.
However, only GPT-2 has been employed in this work, although GPT-3 achieves higher performance in various tasks. 
As stated in the paper, the main reason is GPT-3's difficulty in engineering prompts to perform the expected generation task.
Wang et al.~\cite{DBLP:journals/corr/abs-2303-07678}, proposed query2doc, a query expansion approach that can improve sparse and dense retrieval systems.
It leverages \textit{text-davinci-003} to generate pseudo-documents using few-shot prompting and expand the query with the generated pseudo-documents, similarly to~\cite{DBLP:conf/webi/Claveau21}.
Experimental results show that query2doc improves the performance of BM25 by 3\% to 15\% on ad-hoc IR datasets, such as MS-MARCO, without any model fine-tuning. Additionally, the method benefits state-of-the-art dense retrievers in terms of both in-domain and out-of-domain results.
In~\cite{DBLP:journals/corr/abs-2303-08652}, the authors analyze various conditional text generation techniques and compare their performance to rule-based baselines, aiming to understand whether one can automatically formulate search queries based on factual statements that are similar to those formulated by human experts.
To that aim, they introduce a dataset for fact-checking and evidence collection. 
They establish that similarity to human-created search queries is a valuable indicator of the effectiveness of automatically generated queries in retrieving the same evidence. 
However, they also note that there can be cases where seemingly different search queries may result in collecting the same evidence.

Lee et al.~\cite{DBLP:conf/iui/LeeCJJ23} utilize an LLM and the text from titles and abstracts of research papers to generate keywords for a research paper.
Their analysis, suggests that an LLM has the capability to automatically generate keywords, showcasing its potential in this task.
Lastly, the work of Wang et al.~\cite{DBLP:journals/corr/abs-2302-03495} has several commonalities with ours.
In detail, the authors also leverage ChatGPT and instruct it to create boolean queries that enhance retrieval effectiveness. 
They focus on the task of systematic literature review aiming at retrieving studies related to the review topic.
Similar to our work, they also develop various prompts with increasing complexity, including prompts containing example Boolean queries and guided prompts.
Their prompts are designed for two tasks, i.e. query generation and refinement. 
Their findings showed that when ChatGPT instructed to include MeSH terms for some queries, those MeSH terms were incorrect. 
Besides this limitation, their empirical evaluation suggests that the generated queries result in high search precision, although at the expense of recall.
Also, this study highlighted the ability of ChatGPT to comprehend detailed instructions and create queries with a high level of accuracy, particularly in cases where time is limited and a compromise between precision and recall is acceptable, which is exactly the case in the task studied in our work, i.e. clinical trials retrieval.
\subsection{Retrieving Clinical Trials}
As clinical trials retrieval, we refer to the retrieval task in which, given an admission note that contains patient-related information (query), the search engine aims to retrieve clinical trials (document collection) in which the patient can participate (i.e. retrieve eligible clinical trials).

In the literature, several retrieval approaches have been proposed to address the task of retrieving eligible clinical trials~\cite{DBLP:conf/sigir/KoopmanZ16,DBLP:conf/sigir/AgostiN019,DBLP:journals/jbi/RybinskiXK20,DBLP:conf/sigir/RybinskiKK21}. 
However, particular research attention has been given to this retrieval task through the TREC initiative and specifically the TREC Clinical Trials 2021\footnote{\href{http://www.trec-cds.org/2021.html}{TREC Clinical Trials 2021 Track, accessed on 20/04/2023.}} and 2022\footnote{\href{http://www.trec-cds.org/2022.html}{TREC Clinical Trials 2022 Track, accessed on 20/04/2023.}}tracks.
In both of these tracks, the queries are written in free text and contain patient-related information, such as family history and the patient's past medical history.
This text format imposes many obstacles during retrieval, as outlined in~\cite{sivarajkumar2023clinical}.
The study of Peikos et al.~\cite{DBLP:conf/ecir/PeikosAPV23}, used various patient-related information to construct distinct queries and investigate their effect on retrieving clinical trials.
Specifically, their findings highlight that certain query representations positively affect retrieval effectiveness for patient allocation in clinical trials, especially those that remove patient information related to family history.
Moreover, various research studies that have been published within the TREC Clinical Trials 2021 \footnote{\href{https://trec.nist.gov/pubs/trec30/trec2021.html}{TREC 2021 Publications, accessed on 15/04/2023.}} and 2022 tracks\footnote{\href{https://trec.nist.gov/pubs/trec31/index.html}{TREC 2021 Publications, accessed on 15/04/2023.}}, employ some unsupervised query pre-processing or expansion techniques, such as KeyBERT~\cite{grootendorst2020keybert}.
Also, the proposed systems in several works extract conditions, medical procedures or drugs related to a patient and expand them using, for instance, the Unified Medical Language System\footnote{\href{https://uts.nlm.nih.gov/uts/umls/home}{UMLS Metathesaurus Browser, accessed on 31/04/2023.}}. 

Our detailed analysis of TREC's publications has highlighted some common practices among the participating teams.
It has been found that most of the submitted works filter out (i.e. remove from the final ranking) those clinical trials for which the patient does not meet the required demographic constraints (gender and age). 
However, two additional recruitment conditions, i.e. recruitment status (clinical trials recruitment phase has a specified time window) and location (many trials enrol patients at specific locations), have not been considered in TREC initiative~\cite{soboroff2021overview}.
As a result, the submitted works did not consider these aspects.

The top-performing retrieval approach in TREC 2021~\cite{10.1145/3477495.3531853} relies on a multi-stage retrieval setting that consists of an initial neural query synthesis step that leads to forty distinct query representations.
Those representations are used for retrieval, and the obtained document rankings are fused.
Finally, a two-stage neural re-ranking pipeline trained on clinical trial matching is exploited to create the final ranking.
The most successful approach in the TREC 2022 Clinical Trials track, namely frocchio\_monot5\_e by team h2oloo, employs the Mono-T5 model to re-rank an initially retrieved set of clinical trials\footnote{\href{https://trec.nist.gov/pubs/trec31/papers/Overview_trials.pdf}{Overview of the TREC 2022 Clinical Trials Track, accessed on 31/03/2023.}}.
However, further information about the experimental details of this approach has not been provided.
Notably, the TREC 2021 SoA approach relies on an initial retrieval step that employs the BM25 model with neural query generation and query expansion; it is also plausible that this is the case for the TREC 2022 approach as it has been submitted by the same team and exploits the same model.
Lastly, Liu et al. \cite{DBLP:journals/jamia/LiuYBCLTW19} leverage NLP techniques to enhance retrieval. In detail, the authors created an IR system that initiates a question-answering interactive session with its end-user to eliminate those trials for which the considered patient is explicitly excluded. 
However, even in this interactive IR system, an initial retrieval step is necessary to reduce the number of the considered clinical trials. 
Generally, due to the task's complexity, the proposed approaches in the literature often leverage NLP methods to enhance retrieval performance.

Over the years, several studies have conducted a systematic analysis of the literature around medical and clinical IR~\cite{DBLP:journals/csur/TamineG22,hersh2020information,10.1007/978-3-319-63673-3_65}, including clinical trials retrieval~\cite{sivarajkumar2023clinical}. 
Here, we briefly mention their scope, starting from works that offer a broad overview of health informatics, and concluding with those that investigate a specific research area. 
To begin with, William Hersh, in the fourth edition of his book entitled ``Information Retrieval: A Biomedical and Health Perspective,'' overviews IR systems under the scope of bio-medicine and health domains~\cite{hersh2020information}.
Another work~\cite{10.1007/978-3-319-63673-3_65} analyzes publications and tools by focusing on the diversity of possible medical users and common issues, such as the diverse user vocabulary.
Tamine and Goeuriot~\cite{DBLP:journals/csur/TamineG22} conducted a literature review of semantic IR in the medical domain.
After introducing the medical domain and the available data sources, the authors present an overview of the employed models and techniques before concluding their work by presenting open challenges and future research directions.
Lastly, Sivarajkumar et al.~\cite{sivarajkumar2023clinical} focused on clinical IR, particularly methods, tools, and techniques that leverage free-text electronic health records. 
The authors included 184 research works published from 2012 to 2023 in their analysis.
Their findings show that despite recent technological advancements, a significant amount of clinical IR systems rely on the BM25 model due to its rapid retrieval capability.

In summary, this section provided a comprehensive review of relevant research studies, emphasizing the shift from basic prompting to in-context learning. We explored prominent prompting techniques proposed in the literature and discussed the limitations and drawbacks of LLMs. Additionally, we specifically examined the utilization of LLMs in medical information extraction tasks and their application in the field of information retrieval (IR). The section concluded by presenting relevant studies pertaining to clinical trial retrieval.
The subsequent section describes our implementation of ChatGPT as an intermediate step within a retrieval pipeline focused on clinical trial retrieval.
\section{Using ChatGPT for Medical Information Extraction}
\label{sec:our_approach}
Before the development of LLMs, information extraction from clinical notes was commonly performed by various domain-specific pre-trained models fine-tuned to extract essential pieces of patient-related information~\cite{landolsi2023information}.
However, particular information extraction tasks, such as extracting a patient's lifestyle factors, can not be performed due to a lack of available models~\cite{DBLP:conf/ecir/PeikosAPV23}.
Therefore, due to its ability to comprehend text, ChatGPT might be a plausible alternative.
In this work, ChatGPT is utilized as an intermediary tool to extract information from clinical notes and generate queries that enhance clinical trials retrieval.
There are several reasons behind our decision to employ this model instead of other LLMs such as GatorTron or Med-PaLM~\cite{DBLP:journals/corr/abs-2212-13138}.
To begin with, due to the popularity of ChatGPT compared to other domain-specific LLMs, a vast amount of information about prompt engineering in this model is publicly available\footnote{\href{https://github.com/f/awesome-chatgpt-prompts}{Awesome ChatGPT Prompts, accessed on 21/04/2023.}}$^{,\thinspace}$\footnote{\href{https://github.com/openai/openai-cookbook/blob/main/techniques_to_improve_reliability.md}{Techniques to improve LLM's reliability
, accessed on 21/04/2023.}}.
The availability of these resources, along with the significant amount of research works related to ChatGPT, provide essential insights that aid our investigation. 
In addition, in this work, we mainly focus on the simple task of medical IE and not on more complex NLP tasks, such as medical question answering or medical reasoning.
As we have already mentioned, the GatorTron model has been found to perform marginally better than previous SoA models in medical IE tasks~\cite{DBLP:journals/npjdm/0015CPSSPC0CFZM22}.
Lastly, our study explores some prompting approaches that mimic a user-system conversation; we refer to these as two-step processes.
Therefore, ChatGPT is more suitable to be employed in this context.
The aforementioned ChatGPT's ability, i.e. user-system conversations, makes it also more suitable than \textit{text-davinci-003} model.
In addition to that, it is also cost-effective while it provides comparable performance to \textit{text-davinci-003}, as it has a lower price per token\footnote{\href{https://platform.openai.com/docs/guides/chat}{OpenAI Guides, accessed on 21/04/2023.}}.

We also acknowledge the limitations of LLMs and we consider them in our implementation.
As outlined in Sections \ref{sec:intro} and \ref{sec:lit_rev} the consideration of possible shortcomings is of great importance in the medical domain.
Some LLMs drawbacks are related to their response consistency, generation of hallucinated content, among others.
Regarding the hallucination effect, we argue that is of minor importance in the consider task, as the model is instructed to extract, modify and structure information, rather generate or make decisions.
Also, by employing the proposed approach as an intermediate step in a retrieval pipeline, any potential system failures would only impact the effectiveness of trial enrollment rather than posing direct harm to patients.
Finally, to handle the consistency of the generated responses we relied on the dedicated ChatGPT parameters, as described below.

ChatGPT has several parameters that influence the model's generated response\footnote{\href{https://platform.openai.com/docs/api-reference/chat/create}{OpenAI API, accessed on 21/04/2023.}}, namely \textit{temperature}, \textit{top\_p}, \textit{n}, \textit{stream}, \textit{stop}, \textit{max\_tokens}, \textit{presence\_penalty}, \textit{frequency\_penalty}, and \textit{logit\_bias}.
Also, using the API, it is possible to alter the system's role; this possibility has been exploited by a limited number of works in the literature~\cite{DBLP:journals/corr/abs-2304-05335,DBLP:journals/corr/abs-2303-11568}.
In our experiments, we have carefully set the system's role either using the specific variable that can be set through the API or by specifying it in the prompt text. 
The definition of the system's role has a direct impact on the provided responses, influencing both the content generated and the level of expertise reflected in those responses.
It also imposes ethical boundaries for the generated responses. In some cases, the generated content can be significantly altered due to the selected system role, as recent empirical findings suggest~\cite{DBLP:journals/corr/abs-2304-05335}.
Considering an example, a system role assigned as ``friendly assistant'' might use more casual language in the response. In contrast, a role defined as a ``professional consultant'' might lead the system to generate responses in more formal language. 
This study aims to investigate a high-level distinction, as we set the system's role to either a \textit{general assistant} or a \textit{medical assistant}.
According to the American Association of Medical Assistants (AAMA), medical assistants are involved in various administrative duties, among which is that of updating and filing patient medical records\footnote{\href{https://www.aama-ntl.org/medical-assisting}{American Association of Medical Assistants, accessed on 18/04/2023.}}.
Therefore, in our experiments, the system's roles have been set as \textit{medical assistant}; Qiu et al. also set the system's role as \textit{medical assistant}~\cite{DBLP:journals/corr/abs-2303-11568}.

Among the remaining parameters of ChatGPT, in our experiments, we modified \textit{temperature}, \textit{presence\_penalty}, and \textit{frequency\_penalty} parameters according to the needs of the considered experiment, as we will explicitly comment on the following sections.
Based on the documentation, \textit{temperature} and \textit{top\_p} parameters should not be altered together, as they control the robustness of the model's response.
Higher \textit{temperature} values like 0.8 make the output more random, while lower values like 0.2 will make it more focused and deterministic. However, even a \textit{temperature} value of zero may lead to small variability in the generated responses.
Regarding the \textit{presence\_penalty}, and \textit{frequency\_penalty} these have a range between $[-2,2]$.
Positive \textit{presence\_penalty} values penalize new tokens based on whether they appear in the text so far.
As a result, the model is more likely to add new topics in the generated response.
Positive \textit{frequency\_penalty} values decrease the model's ability to repeat the same tokens.

The designed prompts in our work can be classified into three categories based on their purpose. 
Those that guide ChatGPT to create queries for clinical trials retrieval (Section~\ref{sec:crea_quer}), those that extract specific information from clinical notes (Sections~\ref{sec:key_extr}~and~\ref{sec:bert_chat_section}), and those that identify medical entities and disambiguate their meaning (Section~\ref{sec:ie_meaning}).
Based on the assigned system role through the dedicated variable, the employed prompts are divided into those in which the system acts like a medical professional (Sections~\ref{sec:QGMT}, \ref{sec:domspec_extraction}, \ref{sec:ie_meaning}, \ref{sec:bert_chat_section}) and those in which the system has a generic role (Sections~\ref{sec:QGGT}, \ref{sec:general_extraction}).
In one prompt, described in Section~\ref{sec:general_extraction}, the system does not have a domain-specific role nor knowledge about the general task to be performed, i.e. clinical trials retrieval. 
All employed prompts are zero-shot except the one presented in Section~\ref{sec:bert_chat_section}, which is a two-shot prompt.
In most prompts (except the one presented in Section~\ref{sec:general_extraction}), the model has been discouraged from elaborating or reasoning upon its response.
Lastly, the model is instructed to provide its answer in a specific format in all prompts.

The following subsections, particularly Tables~\ref{tab:query_generation} to \ref{tab:chat_bert_table}, comprehensively describe the prompts used in our study.
In each section, we comment on the motivation behind the usage of the particular prompt and the selected ChatGPT parameters. 
Moreover, to demonstrate the behaviour of ChatGPT, we present the models reply to the employed prompts when the following clinical note is used as input replacing the \textit{{Clinical Note}} token when mentioned:
\begin{quote}
    \textit{``Patient is a 45-year-old man with a history of anaplastic astrocytoma of the spine complicated by severe lower extremity weakness and urinary retention s/p Foley catheter, high-dose steroids, hypertension, and chronic pain. The tumor is located in the T-L spine, unresectable anaplastic astrocytoma s/p radiation. Complicated by progressive lower extremity weakness and urinary retention. The patient initially presented with RLE weakness where his right knee gave out with difficulty walking and right anterior thigh numbness. MRI showed a spinal cord conus mass which was biopsied and found to be anaplastic astrocytoma. Therapy included field radiation t10-l1 followed by 11 cycles of temozolomide 7 days on and 7 days off. This was followed by CPT-11 Weekly x4 with Avastin Q2 weeks/ 2 weeks rest and repeat cycle.''}
\end{quote}

Each sub-section concludes by acknowledging the identified limitations and issues encountered in the obtained ChatGPT generated responses.
Lastly, we discuss alternative prompt approaches that were employed but did not meet the anticipated performance nor behaviour, aiming to provide valuable insights to the research community.
These specific prompts will be included in dedicated Appendices (\ref{sec:appenda}, \ref{sec:appendb}).
\subsection{Guide ChatGPT to formulate queries for clinical trials retrieval}
\label{sec:crea_quer}
In various research works related to clinical trials retrieval, a straightforward and simple practice is to use the raw clinical note (patient's information written in free text) as input in a search engine to retrieve clinical trials for which the patient can be eligible.
Motivated by this, the first step of our investigation on using ChatGPT as an intimidating step to improve retrieval performance in this task focuses on prompts that directly instruct it to generate queries.
Similarly to us, also Wang et al.~\cite{DBLP:journals/corr/abs-2302-03495} have instructed ChatGPT to directly construct boolean queries for the task of retrieving studies for a given systematic literature review. 

In our research, we conducted experiments using various prompting approaches with the goal of explicitly generating search queries.
We encountered some approaches that did not lead to satisfactory retrieval results and others that generate plausible hallucinated responses.
These approaches are presented in~\ref{sec:appenda}, as they might offer further insights to the community.
Specifically, we discuss a few-shot approach and one that generates possible hallucinations.
However, in this section, our focus is on analyzing the two types of prompting approaches that demonstrated the highest retrieval performance when it comes to query creation.
The first approach presented in Section~\ref{sec:QGMT}, incorporates domain-specificity by setting the system's role and also provides a task description in the system's role and prompt text. 
The requested system's output is a set of keywords to be used for clinical trial search.
In the second approach, described in Section~\ref{sec:QGGT}, the system's role is generic, meaning that the model has no knowledge about the medical domain.
However, this is becoming explicit through the employed prompt text in which both the task is described but also, the system's role is defined in the first sentence.
Another difference between the two prompting approaches is related to the details included in the instructions (i.e. the prompt text).
In conclusion, while both tables serve the purpose of guiding the LLM-based system in generating keyword-based queries for clinical trials retrieval, the first table offers more clarity and task-specific context.

\subsubsection{Query generation with domain-specific system role and provided task description}
\label{sec:QGMT}
In the prompt presented in Table~\ref{tab:query_generation}, ChatGPT serves as a helpful medical assistant, with its primary role being to retrieve eligible clinical trials for a medical patient using the provided medical note.
The nature of the task, specifically retrieving clinical trials, is clearly stated in the prompt text and also within the defined system's role.
Within this setting, the model generates a single keyword-based query as its output; an example response to the selected topic is presented in the final column.
\begin{table}[!htbp]
\setlength\tabcolsep{0.5em}
\begin{tabular}{p{5cm}|p{5cm}|p{5cm}}
  \multicolumn{1}{c|}{\centering System's Role} & 
  \multicolumn{1}{c|}{\centering Prompt Text} &
  \multicolumn{1}{c}{\centering Model's output (example)} 
\\ \hline
You are a helpful medical assistant that needs to retrieve eligible clinical trials for your medical patient. You can be synthetic and include all possible topics that you believe this patient can participate based on the provided medical note. &
From the following medical note, use all available information and also related medical concepts, symptoms, diseases, synonyms, or other information to create a keyword-based query that can be used to retrieve clinical trials. Do not include the same keywords twice. Respond with a comma-separated list of keywords that will be used for search. Do not explain or elaborate.
\newline
Medical note: \textit{\{Clinical Note\}}
&
``anaplastic astrocytoma'', astrocytoma, pilocytic, spinal, ``cord tumor'', ``urinary retention'', ``bladder dysfunction'', ``lower extremity weakness'', ``leg weakness'', paraesthesia, ``radiation therapy'', surgery, temozolomide, hypertension, ``chronic pain''
\end{tabular}
\caption{Prompt designed to generate queries for clinical trials retrieval. The table presents the assigned system's role, input prompt, and an example response. In this prompt, the system's role contains  a detailed description of the user and the task.}
\label{tab:query_generation}
\end{table}
In this setting, our goal is to provide the model with a lot of contextual information about its role and the goal we would like it to accomplish. Our intention also is to give it freedom using phrases like \textit{``You can be synthetic and include all possible topics that you believe this patient can participate based on the provided medical note.''}
In addition, we set the \textit{temperature} parameter equal to 0 to ensure that the generated response can be as deterministic as possible.
The \textit{frequency\_penalty} is equal to 1.5 and the model is instructed through the prompt text to avoid having multiple occurrences of the same terms.
Finally, we set the \textit{presence\_penalty} equal to 1, so that the model might have more flexibility to include new terms.
Due to the simplicity of the requested output, i.e. a comma-separated list of keywords, the generated responses were consistent and there was no need to curate them.
As it can be seen in Table~\ref{tab:query_generation}, some of the extracted information is in the form of bi-grams or even tri-grams.
Some of them are explicitly mentioned in the provided clinical note (e.g. \textit{anaplastic astrocytoma}),
while others have been probably inferred from the content (e.g. \textit{cord tumor}).
Also, we observed that the generated response contains some new terms like \textit{pilocytic} or \textit{surgery}, probably generated due to the selected model parameters and the prompt instructions.
However, in this qualitative example, one can observe that various patient-related information that might be important for the retrieval task has not been extracted (e.g. \textit{high-dose steroids, CPT-11, Avastin}).
\subsubsection{Query generation with generic system role and provided task description}
\label{sec:QGGT}
In the prompt presented in Table~\ref{tab:query_gen_refine}, the system acts as a helpful assistant, but its role is later clarified through the input prompt.
The task is explicitly mentioned and contains more specific information about the collection from which the clinical trials will be retrieved.
Here, the model is instructed first to create a keyword-based query based on information present in the inputted clinical note, and then (in the same prompt), it is instructed to refine it based on synonyms or related medical concepts.
Both of the requested outputs are created in a single interaction with the system.
The model's parameters are identical to those of the previous prompt, as the prompt serves the same purpose (\textit{temperature} is 0, \textit{frequency\_penalty} is 1.5, and  \textit{presence\_penalty} is 1).
\begin{table}[!htbp]
\begin{tabular}{p{9cm}|p{6cm}}
  \multicolumn{1}{c|}{\centering Prompt Text} &
  \multicolumn{1}{c}{\centering Model's output (example)} 
  \\ \hline
Act as a medical assistant. Your task is to retrieve clinical trials from a registry of clinical trials in the United States. To achieve that, you have access to a medical clinical note of a patient.
\newline\newline
Follow my instructions precisely to extract the requested information from a patient’s medical clinical note. Do not explain or elaborate. Respond with exactly what I request, and reply in the requested format.
\newline\newline
1. Write a keyword-based query that can be used in a search engine to search for clinical trials in which this patient can participate. {[}query\_keywords{]}
\newline
Answers' format: {[}query\_keywords{]} ``query\_text'' {[}query\_keywords{]}
\newline\newline
2. Refine the query based on further details, such as synonyms or related medical concepts. [query\_keywords\_expanded] 
\newline
Answers' format: {[}query\_keywords\_expanded{]} ``query\_text'' {[}query\_keywords\_expanded{]}
\newline\newline
Do not mention the terms clinical trial in the created queries as we search in a collection of clinical trials.  Do not use abbreviations, use the resolved abbreviation format.
\newline\newline
Medical note: \textit{\{Clinical Note\}} &
[query\_keywords] ``clinical trial anaplastic astrocytoma spine radiation temozolomide Avastin CPT-11'' [query\_keywords]
 \newline\newline
[query\_keywords\_expanded] ``clinical trial anaplastic astrocytoma spine radiation temozolomide Avastin CPT-11 neuropathic pain lower extremity weakness urinary retention hypertension'' [query\_keywords\_expanded]
\end{tabular}
\caption{Prompt aiming at query generation, with an explicit mention about the task to be performed and the system's role in the prompt text.
The system's Role is generic, i.e. ``You are a helpful assistant.''}
\label{tab:query_gen_refine}
\end{table}
Probably due to the prompt's complexity, this setting has two empirically identified shortcomings.
First, the system fails to follow the instructions to avoid the term \textit{clinical trial}, and the usage of abbreviations in the generated response.
Second, for some topics, the \textit{``[query\_keywords]''} or  \textit{``[query\_keywords\_expanded]''} tokens were missing from the end of the generated responses.

Regarding the quality of the generated response in the demonstrated example, one can observe that the first query contains the patient's medical condition, and then the patient's medication and therapy.
When the model is asked to refine the query, it successfully retained the original terms and expanded them with extra terms, including the term \textit{neuropathic}, that was not present in the original clinical note.
This extraction pattern and query creation is observed for the vast majority of the queries in the used collections.
However, as one can observe based on the example model's output, various patient-related information that might be essential for this task is missing. 
\subsection{Guide ChatGPT to perform information extraction}
\label{sec:key_extr}
We investigated the capability of ChatGPT to extract particular information from a provided clinical note, in several settings, two of which have competitive performance and are described in Sections ~\ref{sec:general_extraction} and~\ref{sec:domspec_extraction}.

Specifically, Section~\ref{sec:general_extraction} describes a two-step approach (two interactions with the system) that utilizes ChatGPT as a general-purpose helpful assistant, aiming at extracting a list of keywords that fully describe the content of a given text (step 1) and then expand them with related keywords (step 2). 
Consequently, the model outputs two distinct lists of keywords.
In this approach, the system's role is generic, while we have not specified the type of information to be extracted nor the actual task to be performed.
The second approach presented in Section~\ref{sec:domspec_extraction} sets ChatGPT's role as a medical assistant, focusing on identifying medical conditions, treatments, and related terminology. This approach demands knowledge of medical terminology and abbreviations. 
Also, in this approach, information extraction and expansion have been instructed in a single step, and the purpose (i.e. search for clinical trials) is mentioned in the prompt (``identify clinical trials of interest'').

The main commonality between these settings is their objective to provide a condensed representation of the content in the form of keywords, which can be used for searching eligible clinical trials.
The main differences between the two approaches are related to the scope and domain knowledge requirements (i.e. allocated through the selected system's role), and in their complexity (i.e. multi-step and single step information extraction and expansion).

Lastly, in~\ref{sec:appendb} we present a prompting approach that explicitly guides the keyword expansion process.
Despite the fact that the presented approach did not outperformed those presented here, it provides some valuable insights regrading the usage of ChatGPT for query expansion.
\subsubsection{Two-step generic information extraction and expansion}
\label{sec:general_extraction}
In this setting, ChatGPT is completely task and domain agnostic.
The model's \textit{temperature} is equal to zero, the \textit{frequency\_penalty} is set to 2 (maximum value) and the \textit{presence\_penalty} to 1.
We have maximized the \textit{frequency\_penalty} because the prompt does not contain any instructions to avoid same term generation.
Here, as we see in Table~\ref{tab:keyword_extraction_nodomain}, the model outputs two sets of keywords (one per system interaction), the first contains extracted terms (in form of uni-grams, bi-grams, etc.) and the second that expands them with 20 additional related terms.
The selection of the number of expansion terms is motivated by traditional pseudo-relevance feedback approaches that request a specific number of related terms to be added in the original query.
By combining the model's outputs we run three distinct retrieval searches; one using the extracted terms, one only with the expanded, and one with their combination.
\begin{table}[!htbp]
\setlength\tabcolsep{0.5em}
\begin{tabular}{p{5cm}|p{11cm}}
\multicolumn{1}{c|}{\centering Prompt Text} &
\multicolumn{1}{c}{\centering Model's output (example)} 
\\  \hline
From the following text, extract a list of keywords that fully describe its content.
\newline\newline
text:\textit{\{Clinical Note\}} &
``45-year-old man'', ``anaplastic astrocytoma'', spine, ``lower extremity weakness'', ``urinary retention'', ``Foley catheter'', ``high-dose steroids'', hypertension, ``chronic pain'', ``T-L spine'', radiation, ``progressive lower extremity weakness'', ``RLE weakness'', ``right knee'', ``right anterior thigh numbness'', ``spinal cord conus mass'', biopsied, ``field radiation'', t10-l1, temozolomide, CPT-11, Avastin \\ \hline
Expand the extracted list of keywords with 20 additional related keywords. &
``progressive weakness'', unresectable, cycles, tumor, chemotherapy, ``neurologic examination'', neurosurgery, oncology, ``palliative care'', ``imaging studies'', radiotherapy, corticosteroids, ``nerve compression'', ``muscle weakness'', ``bladder dysfunction'', ``motor deficits'', paraplegia, ``spinal cord'',``cancer treatment''
\end{tabular}
\caption{Two-step process (one input-output per row) of Information Extraction from a clinical note text. In the first step, we instruct ChatGPT to extract the information, while in the second-step, we instruct it to expand these keywords.
System's role:``You are a helpful assistant.''}
\label{tab:keyword_extraction_nodomain}
\end{table}
Regarding the generated responses, as it can be seen in the first row of Table~\ref{tab:keyword_extraction_nodomain}, the extracted keywords and phrases indeed describe the information of the clinical note almost completely.
Specifically, for the example query, only the bi-gram \textit{``difficulty walking''} has not been extracted.
This behavior has been observed for the majority of the queries in the employed collections. 
In the second interaction with the system, ChatGPT is instructed to expand the list of extracted keyword with related keywords.
By observing the generated keywords (e.g. \textit{``palliative care''} or \textit{``imaging studies''}) one can conclude that some of the selected terms are describing broader medical concepts. 
As a result, it is expected that using these terms for retrieval might lead to decrease in precision-oriented measures.
\subsubsection{Single-step domain-specific information extraction and expansion}
\label{sec:domspec_extraction}
In this setting, domain specificity has been achieved by setting the system's role and by mentioning the search task.
The model's \textit{temperature} is equal to zero, the \textit{frequency\_penalty} is set to 2 (maximum value) and the \textit{presence\_penalty} to 1.
However, in this setting ChatGPT is instructed to extract only information related to the patient's medical condition and treatments.
\begin{table}[!htbp]
\setlength\tabcolsep{0.5em}
\begin{tabular}{p{8cm}|p{8cm}}
\multicolumn{1}{c|}{\centering Prompt Text} &
\multicolumn{1}{c}{\centering Model's output (example)} 
\\ \hline
Please identify the patient's medical condition and current treatments, including any alternative names, abbreviations, or synonyms for these terms, as well as any additional criteria that may be important for identifying clinical trials of interest.
Respond with a comma-separated list of keywords that will be used for search. Do not elaborate or explain. 
\newline\newline
Patient's medical note: \textit{\{Clinical Note\}} &
``anaplastic astrocytoma'', spine, ``lower extremity weakness'', ``urinary retention'', ``Foley catheter'', ``high-dose steroids'', hypertension, ``chronic pain'', ``T-L spine'', radiation, ``RLE weakness'',  temozolomide, ``CPT-11 Weekly'', Avastin 
\end{tabular}
\caption{Single-interaction setting in which ChatGPT is instructed to extract only information related to the patient's medical condition and treatments.
System's role:``You are a helpful medical assistant.''}
\label{tab:keyword_extraction_focused}
\end{table}
The selection of these particular entities is inspired by a related study~\cite{DBLP:conf/ecir/PeikosAPV23}, in which the findings suggested that using a patient's medical problem, treatments and tests as query improves performance in clinical trials retrieval.
The generated responses were consistent in terms of their format. 
However, although the model has been instructed to include alternative names, abbreviations, or synonyms, for the vast majority of queries, its responses only contain extracted keywords.
\subsection{Guide ChatGPT to perform information extraction and entity meaning disambiguation}
\label{sec:ie_meaning}
A shared disadvantage among the previously introduced prompting settings is their inability to disambiguate the meaning of the extracted keywords and phrases.
To overcome that, we investigated ChatGPT's ability to summarise the patient's medical information by resolving abbreviations, recognizing key information, and adding pertinent MeSH terms.
As a result, we have designed the prompt presented in Table~\ref{tab:entities_ent_dissambiquation}, that aims to extract information that is most commonly found within a patient's clinical note.
In addition, by employing this prompt, one can possible extract a patient's lifestyle factors, a piece of information for which, there is not a publicly available model to extract it~\cite{DBLP:conf/ecir/PeikosAPV23}.
The model's \textit{temperature} is equal to zero, and both the \textit{frequency\_penalty} and the \textit{presence\_penalty} are set to 0.
These values have been selected to allow the model to use the same terms if needed, for instance when a diagnosis and a medical problem are identical.
\begin{table}[!htbp]
\setlength\tabcolsep{0.5em}
\begin{tabular}{p{8cm}|p{8cm}}
  \multicolumn{1}{c|}{\centering Prompt Text} &
  \multicolumn{1}{c}{\centering Model's output (example)} 
  \\ \hline
Follow my instructions precisely to extract the requested information from a patient’s medical clinical note. Do not explain or elaborate. Respond with exactly what I request, and reply in the requested format.
\newline\newline
From the following clinical note, resolve all the abbreviation mentioned in the text, and then extract: the patient’s, age, gender, medical problem, diagnosis, diseases, symptoms, medications, drugs, dosages, treatments, medical history, family history, lifestyle factors, lab examinations, lab results, vital signs. Add MeSH terms that are relevant to the patient’s medical problem, diagnosis or disease. If you are unable to extract the information, write 'N-A'.
Answer in JSON format: \newline\newline
\{
    ``answer'':\{ \newline
        ``abbreviations'': ``resolved abbreviations'', \newline
        ``diagnosis'':``'', 
        ``medical\_problem'':``'',\newline
        ``diseases'':``'',
        ``drug'':``'',
        ``dosages'':``'',
        ``symptoms'':``'',
        ``treatments'':``'', \newline
        ``medications'':``'',
        ``medical\_history'':``'',\newline
        ``family\_history'':``'',
        ``lifestyle\_factors'':``'',\newline
        ``lab\_examinations'':``'',\newline
        ``lab\_results'':``'',
        ``vital\_signs'':``'',\newline
        ``gender'':``'',\newline
        ``age'':``'',
        ``MeSH\_terms'':``''
    \}\}\newline\newline
     Medical note: \textit{\{Clinical Note\}}
     &
\{
    ``answer'': \{\newline
		``abbreviations'': ``s-p: status post, RLE: right lower extremity, MRI: magnetic resonance imaging, t10-l1: thoracic vertebrae 10 to lumbar vertebrae 1, CPT-11: irinotecan, Q2 weeks: every two weeks'',\newline\newline
		``diagnosis'': ``anaplastic astrocytoma'',\newline
		``medical\_problem'': ``anaplastic astrocytoma of the spine, severe lower extremity weakness and urinary retention'',\newline\newline
		``diseases'': ``hypertension, chronic pain'',\newline
		``drug'': ``steroids, temozolomide, irinotecan, Avastin'',\newline
		``dosages'': ``high-dose steroids, temozolomide 7 days on and 7 days off, CPT-11 Weekly x4 with Avastin Q2 weeks- 2 weeks rest and repeat cycle'',\newline\newline
		``symptons'': ``lower extremity weakness, right anterior thigh numbness, urinary retention'',\newline\newline
		``treatments'': ``field radiation, temozolomide, CPT-11 with Avastin'',\newline
		``medications'': ``steroids, temozolomide, irinotecan, Avastin'',\newline
		``medical\_history'': ``anaplastic astrocytoma of the spine'',\newline
		``family\_history'': ``N-A'',
		``lifestyle\_factors'': ``N-A'',
		``lab\_examinations'': ``N-A'',
		``lab\_results'': ``N-A'',\newline
		``vital\_signs'': ``N-A'',\newline
		``gender'': ``male'',
		``age'': ``45'',\newline\newline
		``MeSH\_terms'': ``Astrocytoma, Spinal Cord Neoplasms, Lower Extremity, Urinary Retention, Steroids, Temozolomide, Irinotecan, Bevacizumab, Radiation, Magnetic Resonance Imaging''
		\}\}
\end{tabular}
\caption{Medical entity extraction and meaning disambiguation. Spaces have been added in the example of the model's output for readability purposes.
System's Role: ''You are a helpful medical assistant that needs to retrieve eligible clinical trials for your medical patient.''}
\label{tab:entities_ent_dissambiquation}
\end{table}
In this prompt setting, both the system's role and the task to be performed are explicitly mentioned.
Then, the prompt text instructs the model to extract patient-related information that can be present in a clinical note.
The model is asked to not elaborate and it was explicitly instructed to not include not answer in case it has limited information.
We added this instruction based on previous findings that suggest that ChatGPT might attempt to infer information~\cite{DBLP:journals/jamia/UzunerSSD11,hu2023zero}.
Similar to a previous work~\cite{DBLP:conf/ecir/PeikosAPV23}, our goal in this setting is to also exploit the various extracted information, synthesize queries in which we (as users) know what entities are included, and use them for clinical trials retrieval.

Generally, this prompt highlights ChatGPT's ability to extract and disambiguate the meaning of several patient-related information. 
However, probably due to the amount of requested information, various minor shortcomings have been identified in our conducted experiments.
Through our experimentation, it has been detected that the formatting of the generated replies (i.e. the exported JSON) is not always consistent (e.g. missing commas), but this is shortcoming is easy to be automatically detected and fixed. 
Also, it is expected that newer GPT versions (e.g. GPT-4~\cite{DBLP:journals/corr/abs-2303-08774}) will follow instructions more accurately.
Other identified limitations were related to the nature of the produced responses.
Occasionally, it was difficult for the model to use the ``N-A'' token, even though it seems that it ``understands'' that it is not capable to fill-in some information. 
In these case, the model's responses contained phrases like \textit{``this information is not explicitly mentioned in the text.''} or \textit{'lifestyle\_factors' :'Not applicable'}. 
Another shortcoming is that the model's response often contained negated content.
For example, \textit{{'lifestyle\_factors': 'non-smoker, non-alcoholic, non-illicit drug user, menopausal'}}, while, for retrieval purposes, ideally, we would prefer only the term ``menopausal'' to be extracted.
As we are not medical experts, we can not conclude about the quality of the generated MeSH terms, however other studies showed that the quality of suggested MeSH terms by ChatGPT is poor~\cite{DBLP:journals/corr/abs-2302-03495}.
Lastly, we observed that not all available information in a clinical note was extracted in the first system interaction.
As a result, we suggest that maybe a multi-step interaction with the system should be preferred instead, or other prompting approaches, like the consistency method~\cite{DBLP:journals/corr/abs-2203-11171}.
This approach will raise the operational cost, but it might lead to better extraction quality.
In our study, we employed this prompt setting due to cost and time constraints.

To conclude, based on this prompting setting, one can observe that, in general, ChatGPT can be used to extract and categorize the various medical information contained in a clinical note.
Also it can be instructed not to infer information that is not present, reducing the problem of hallucination generation.
Therefore, it can be employed to structure the information of clinical notes and aid healthcare personnel by reducing the required time and effort in performing this task.
\subsection{Combine ChatGPT with a transformer-based model for negation removal}
\label{sec:bert_chat_section}
Based on the previous prompting approaches we observed that ChatGPT can identify medical entities with minor errors and can also disambiguate their meaning.
However, as the model is trained to generate responses to questions, it occasionally provides negated content in its answers, especially in the setting presented in the previous section.
That, as mentioned in previous studies, can negatively impact retrieval effectiveness~\cite{DBLP:conf/amia/ChapmanBHCB01,DBLP:conf/sigir/KoopmanZ14,DBLP:conf/ecir/PeikosAPV23}.

\begin{figure}[!htbp]
\includegraphics[width=\textwidth,scale=0.5]{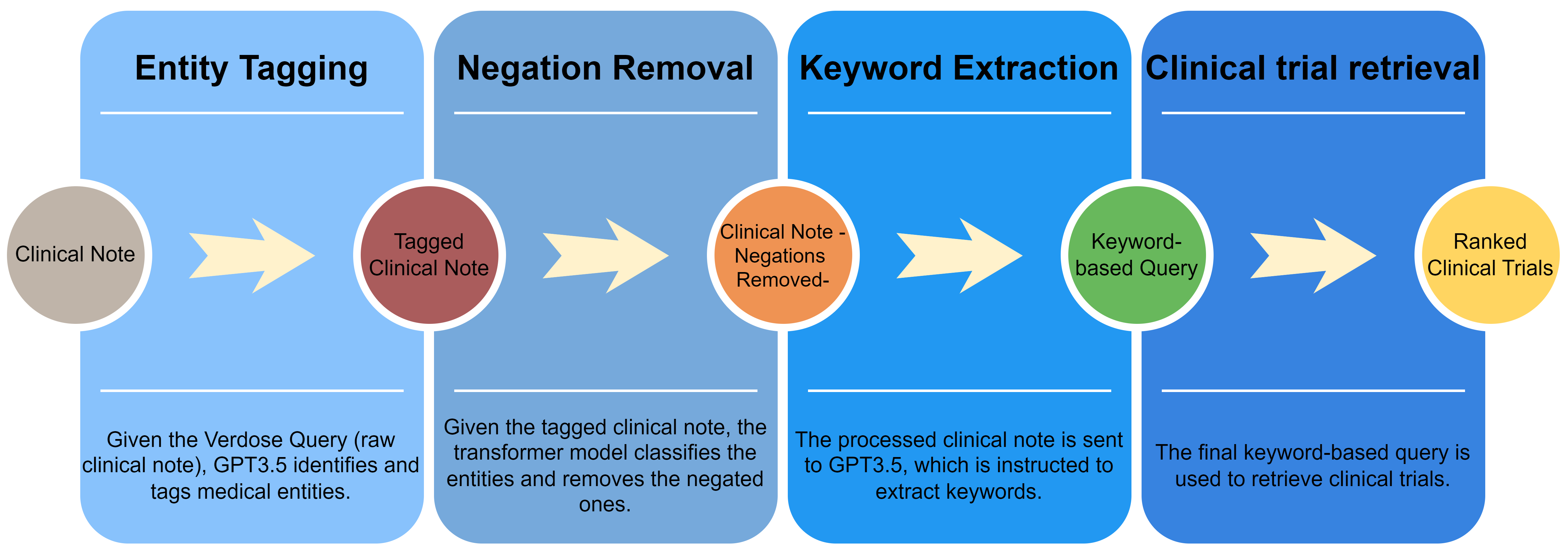}
\caption{Combining ChatGPT with a negation classification model.}
\label{figure:fig1}
\end{figure}

Therefore, this section presents an approach that combines ChatGPT with a pre-trained transformer-based model based on ClinicalBERT~\cite{alsentzer-etal-2019-publicly} that aims to overcome this issue.
The transformer-based model is fine-tuned for medical condition assertion~\cite{van-aken-2021-assertion};
it categorizes medical conditions mentioned in clinical notes as present, absent, or possible.
By doing that, we aim to leverage the power of ChatGPT for medical entity identification and the ability of the pre-trained transformer-based model to identify negations.
Our approach consist of four steps as shown in Figure~\ref{figure:fig1}.
In detail, in the first step a clinical note is given as input to ChatGPT using the prompt text presented in the first row in Table~\ref{tab:chat_bert_table}.
Here, ChatGPT is instructed to rewrite the clinical note by enclosing medical problems, treatments, tests, symptoms, and lifestyle factors within [entity] token.
\textit{Temperature}, \textit{frequency\_penalty}, and  \textit{presence\_penalty} are set to zero, to keep the answer as consistent as possible, and discourage the model to use different terms than those originally present in the clinical note.
However, as it can be seen in the example response, there are entities that the model fails to tag, e.g. \textit{lower extremity weakness}.
To further aid ChatGPT in this task, two examples of the requested annotation are also provided.
By doing that, the clinical note is ready to be used as input to the fine-tuned model introduced in~\cite{van-aken-2021-assertion} in step two.
The model classifies all tagged entities and those found as absent (i.e. negated content) are removed.
In the third step, the ``clean'' clinical note is given as input to ChatGPT through the prompt text presented in the second row in Table~\ref{tab:chat_bert_table}.
This step leverages the keyword extraction approach presented in Section~\ref{sec:domspec_extraction}, with the same model parameters (\textit{temperature} is equal to zero, \textit{frequency\_penalty} is two and the \textit{presence\_penalty} is 1.).
Despite the fact that we employed exactly the same prompt, minor differences in the model's output have been observed.
In this example, the terms ``therapy'' and ``weekly,''are missing, while ``T-L spine'' has been transformed to ``t10-l1 temozolomide.''
\begin{table}[!htbp]
\setlength\tabcolsep{0.5em}
\begin{tabular}{p{8cm}|p{8cm}}
\multicolumn{1}{c|}{\centering Prompt Text} &
\multicolumn{1}{c}{\centering Model's output (example)} 
\\ \hline
Follow my instructions precisely to tag the requested information in a patient’s medical clinical note.  Do not explain or elaborate. Respond with exactly what I request, and reply in the requested format.
\newline\newline
Rewrite the clinical note by adding the token [entity] at the begging and the token [entity] at the end of the mentioned medical problems, symptoms, lifestyle factors, medical treatments and medical tests.
For example,\newline\newline
Input:   The patient recovered during the night and now denies any shortness of breath.
Output:  The patient recovered during the night and now denies any [entity] shortness of breath [entity].
\newline\newline
Input:
An 8-year-old male presents in March to the ER with fever up to 39 C, dyspnea and cough for 2 days
Output:
An 8-year-old male presents in March to the ER with [entity] fever [entity] up to 39 C, [entity] dyspnea [entity] and [entity] cough [entity] for 2 days.
\newline\newline
Do the same for the clinical note: \textit{\{Clinical Note\}}
&
[entity] Patient [entity] is a 45-year-old man with a history of [entity] anaplastic astrocytoma [entity] of the spine complicated by severe lower extremity weakness and urinary retention [entity] s/p Foley catheter [entity], [entity] high-dose steroids [entity], [entity] hypertension [entity], and chronic pain.
The tumor is located in the [entity] T-L spine [entity], [entity] unresectable anaplastic astrocytoma [entity] s/p radiation. Complicated by progressive lower extremity weakness and urinary retention. Patient initially presented with [entity] RLE weakness [entity] where his right knee gave out with difficulty walking and right anterior thigh numbness. MRI showed a spinal cord conus mass which was biopsied and found to be [entity] anaplastic astrocytoma [entity]. Therapy included field radiation [entity] t10-l1 [entity] followed by 11 cycles of temozolomide 7 days on and 7 days off. This was followed by [entity] CPT-11 Weekly x4 [entity] with [entity] Avastin [entity] Q2 weeks/2 weeks rest and repeat cycle.7
\\\hline
Please identify the patient's medical condition and current treatments, including any alternative names, abbreviations, or synonyms for these terms, as well as any additional criteria that may be important for identifying clinical trials of interest. Respond with a comma-separated list of keywords that will be used for search. Do not elaborate or explain. 
\newline\newline
Patient's medical note: \textit{\{Clinical Note\}}
&
``anaplastic astrocytoma'', spine, `lower extremity weakness'', ``urinary retention'', ``Foley catheter'', ``high-dose steroids'', hypertension, ``chronic pain'', ``radiation therapy'',  ``t10-l1 temozolomide'', CPT-11, Avastin
\end{tabular}
\caption{Approach that combines ChatGPT3 with the pre-trained model for negation detection. Each row corresponds to one ChatGPT input-output.
System's role:``You are a helpful medical assistant.''}
\label{tab:chat_bert_table}
\end{table}
Finally, the extracted keywords are used as a query for retrieval, leading to a ranked list of clinical trials.

Prior to our decision to employ this setting to remove negated content from a clinical note, we tried to directly instruct ChatGPT to perform this task.
However, the obtain results did not meet our expectations.
In detail, we experimented with prompts like \textit{``Remove the negated content of the following medical note.
''} or \textit{``re-write the following medical note by removing the negated medical problems, symptoms...''}.
The models responses, were totally altering the meaning of the original clinical note, for instance: 
Given the original text \textit{``He is healthy with no history of allergies...''}, the model identified the negation in the sentence, but its output was \textit{``He is healthy with history of allergies...''}; therefore, we proceeded with the presented approach.
\section{Experiments}
\label{sec:exp_setup}
This section provides detailed information regarding the experimental design and results. In detail, it discusses the employed collections, considered baselines, and specific retrieval configurations utilized during our empirical evaluation. 

To empirically evaluate the quality of the created prompts in terms of retrieval performance, we used the benchmark collections provided in the TREC 2021~\cite{soboroff2021overview} and 2022\footnote{\href{https://trec.nist.gov/pubs/trec31/papers/Overview_trials.pdf}{Overview of the TREC 2022 Clinical Trials Track, accessed on 31/03/2023.}} Clinical Trials tracks.
Hereafter we refer to them as TREC 2021 and TREC 2022, respectively.
The provided queries are patient case descriptions that simulate an admission statement in an EHR.
Their size varies from 5 to 10 sentences, and they have been created by individuals with medical training.
The TREC 2021 collection has a total of 75 queries and the TREC 2022 has 50 queries.
The document collection (common for both tasks) consists of around 376k clinical trials extracted from ClinicalTrials.gov~\footnote{\href{https://clinicaltrials.gov/}{ClinicalTrials.gov, accessed on 25/05/2023.}}, which is  a required registry for clinical trials in the United States.
Unlike the provided queries that are unstructured clinical notes, a clinical trial document follows a structured format that includes sections such as title, studied condition, summary, detailed description, and eligibility section, among others.
All of these sections are important during retrieval, however the eligibility section contains information (i.e. its inclusion/exclusion criteria) that might determine whether a given patient will be eligible to participate in the trial.

Regarding the relevance assessment, a clinical trial has been evaluated as \textit{eligible}, \textit{excludes}, and \textit{not relevant} to a given clinical note, where \textit{eligible} means that the patient can participate in it, \textit{excludes} means that the patient is explicitly excluded, and \textit{not relevant} which means that the patient does not have sufficient information to qualify for the trial.
The employed evaluation measures in both tracks are nDCG@5 and nDCG@10, P@5 (Precision at 5) and P@10, 
R-Precision (Rprec), and mean reciprocal rank (MRR).
Besides the nDCG measure, the remaining measures count only \textit{eligible} trials as relevant, i.e. trials assessed as \textit{excludes} are considered also \textit{not relevant}.
In our evaluation, we follow the official guidelines, and we further report Bpref (binary preference) and P@25. 
For an comprehensive description of the employed evaluation measures, please refer to~\cite{DBLP:journals/ftir/MitraC18} and~\cite{DBLP:books/daglib/0021593}.

For our experiments, we obtained the benchmarks collections (queries and documents) through the~\textit{ir-datasets}~\cite{macavaney:sigir2021-irds}.
To perform the experiments we employed PyTerrier~\cite{pyterrier2020ictir} and index the document collection using the default PyTerrier's parameters, i.e porter-stemming and stopword removal.
For each document, we index all of its available sections.
A summary of the conducted experiments and the selected baselines is presented in Table~\ref{tab:experiments}, where an identifier is assigned to each experiment, followed by a sort description of the query that is used for retrieval and a pointer to the section that describes the employed prompt (when needed).
\begin{table}[!htbp]
\setlength\tabcolsep{0.9em}
\begin{tabular}{>{\centering\arraybackslash}p{6cm}|>{\centering\arraybackslash}p{7cm}|>{\centering\arraybackslash}p{1.5cm}} 
Experiment Details (IDs) & Retrieval Input  & Prompt \\ \hline \noalign{\vskip 0.2em}
BM25 Baseline (\textbf{BM25}) & Original query with the default PyTerrier's pre-processing.  & -  \\
KeyBERT+BM25 Baseline (\textbf{KeyBERT}) & Keywords extracted from the original query using KeyBERT. & -    \\ \hline  \noalign{\vskip 0.2em}    
ChatGPT \textbf{Q}uery \textbf{G}eneration, \textbf{M}edical Role \& \textbf{T}ask Description (\textbf{QGMT}) & A single keyword-based query. & Section~\ref{sec:QGMT} \\ 
ChatGPT \textbf{Q}uery \textbf{G}eneration, \textbf{G}eneric Role \& \textbf{T}ask Description (\textbf{QGGT}) & This prompt outputs two sets of queries. A single keyword-based query, and a refined query that contains novel terms. & Section~\ref{sec:QGGT}  \\\hline \noalign{\vskip 0.2em}    
ChatGPT  \textbf{I}nformation \textbf{E}xtraction, \textbf{G}eneric Role (\textbf{IEG}) & This prompt outputs two sets of queries. One with all extracted keywords and phrases and one with novel expansion terms. & Section~\ref{sec:general_extraction}\\
ChatGPT \textbf{I}nformation \textbf{E}xtraction, \textbf{M}edical Role \& \textbf{T}ask Description (\textbf{IEMT}) & A single keyword-based query. & Section~\ref{sec:domspec_extraction}  \\\hline \noalign{\vskip 0.1em}    
ChatGPT \textbf{I}nformation \textbf{E}xtraction \& Entity \textbf{M}eaning \textbf{D}isambiguation, \textbf{M}edical Role \& \textbf{T}ask Description (\textbf{IEMDMT}) & This prompt outputs various patient-related information. Therefore, many queries variations are constructed and used for retrieval. & Section~\ref{sec:ie_meaning} \\\hline \noalign{\vskip 0.2em}    
\textbf{N}egation \textbf{R}emoval and ChatGPT \textbf{I}nformation \textbf{E}xtraction, \textbf{M}edical Role \& \textbf{T}ask Description (\textbf{NRIEMT}) & A single keyword-based query without negated terms. & Section~\ref{sec:bert_chat_section}   
\end{tabular}
\caption{Summary of the conducted experiments.}
\label{tab:experiments}
\end{table}
As mentioned in Section~\ref{sec:related_work}, the top-performing retrieval approaches in TREC 2021 and TREC 2022 Clinical Trials track rely on the BM25 model for their initial retrieval step.
Similarly, we employed PyTerrier's implementation of the BM25 model, with its default parameters in all retrieval experiments conducted in this work.
Based on that, the first selected baseline exploits the original queries, as provided in the collections, and the BM25 model for retrieval.
We refer to this experiment as \textit{BM25}.
In addition, we leverage KeyBERT to extract keywords from the original queries and use them for retrieval, this experiment is mentioned as \textit{KeyBERT} and it is also serves as a baseline.
We used the original KeyBERT implementation and set its parameters so that it extracts twenty unigrams or bi-grams (\textit{keyphrase\_ngram\_range=(1, 2)}, \textit{top\_n=20}).
Regarding the ChatGPT-based experiments, these are categorized into four categories, based on the purpose of the employed prompt; to those that instruct it to generate queries (\textit{QGMT}, and \textit{QGGT}),
extract information (\textit{IEG} and \textit{IEMT}), extract specific patient-related information (\textit{IEMDMT}), and finally our hybrid approach for negation removal (\textit{NRIEMT}).
In these experiments, the original queries are firstly processed using ChatGPT, with both the general-purposed and task-specific prompts introduced in the previous sections.
We note here that each query has been processed in a distinct system interaction through the API, i.e. new conversation chat.
Then, these generated queries are processed using PyTerrier's default processing steps (porter-stemming and stopword removal) and used to retrieve clinical trials.
ChatGPT leverages the gpt-3.5-turbo	model whose selected parameters have been presented in the previous sections.
Some of the ChatGPT generated queries and the KeyBERT baseline generate n-grams and not just keywords.
In our experiments, we experimented with uni-grams, uni-grams and bi-grams, and uni-grams, bi-grams, and tri-grams for retrieval.
For the employed collections, we found that using only uni-grams, i.e. treating bi-grams and tri-grams as uni-grams lead to better retrieval performance.
Lastly, we use the RM3 model, which is a standard approach for pseudo-relevance feedback query expansion and combine it with all of the aforementioned retrieval pipelines; we exploit its PyTerrier implementation with the number of pseudo-relevance documents equal to ten and twenty expansion terms.

The following sections provide insights related to ChatGPT's ability to create queries for clinical trials retrieval (Section~\ref{sec:res_qg}), extract information from a clinical note to be used for retrieval (Section~\ref{sec:res_ie}).
Section~\ref{sec:res_ecir} show the performance achieved when various patient related information is extracted by ChatGPT used to synthesize queries.
Finally, Section~\ref{sec:res_nr_ie} shows the results obtained when possible negated entities are removed from the queries.
In the presented results in the tables, the statistical significance is tested against the BM25 baseline according to a paired t-test with Bonferroni multiple testing correction, at significance levels 0.05(\ssa).
The best result per measure is in boldface, while the first four rows of each table present the performance achieved by the baseline retrieval pipelines.
\subsection{ChatGPT for query generation}
\label{sec:res_qg}
The results presented in Table~\ref{tab:query_gen_resutls} show the retrieval performance achieved by prompts that instruct ChatGPT to generate a single query (per clinical note) to be used for clinical trials retrieval.

First we comment on the retrieval performance achieved by the BM25 and KeyBERT retrieval pipelines (i.e. the baselines), with and without the incorporation of the RM3 model for query expansion.
One can notice that the performance improvements are not consistent between the two collections.
In the TREC 2021 collection, KeyBERT outperforms BM25 only in terms of Bpref, while in TREC 2022 KeyBERT underperforms only for MRR.
This performance variations might occur due to the retrieval of unjudged documents in the top ranking positions, which are considered as \textit{not relevant} during our evaluation.
Another reason might be the semantic characteristics of the provided queries in the two collections, for instance the number of hard and easy queries.
As hard queries we assume queries for which the patient might be an infant, or the considered disease might be very rare.
However, we have not investigated further towards these directions in our study.
Combining BM25 with the RM3 model (i.e. BM25 + RM3), improves or has a slightly lower performance compared to BM25 across all measures, except MRR where the observed decrease is greater but not statistically significant.
All in all, no single baseline approach consistently outperforms the others across all metrics and collections.
Also, incorporating the RM3 model has mixed effects on the performance of the BM25 and KeyBERT, as it is (probably) affected by the number of relevant documents in the top ten ranking positions. 
\begin{table}[!htbp]
\setlength\tabcolsep{0.5em}
\resizebox{1\textwidth}{!}{%
\begin{tabular}{c|cccccc|cccccc}
                               & \multicolumn{6}{c|}{TREC 2021}                                                   & \multicolumn{6}{c}{TREC 2022}                                                  \\ \hline
                               & Rprec & Bpref & P@10 & P@25 & MRR & nDCG@10   & Rprec & Bpref & P@10 & P@25 & MRR & nDCG@10 \\ \hline
BM25                           & .162        & .184        & .264       & .211       & \textbf{.471}     & .469     & .180        & .172        & .272       & .235       & .507     & .394   \\
BM25 + RM3                       & \textbf{.184}        & .241\ssa    & \textbf{.285}       & \textbf{.231 }      & .463     & \textbf{.484}     & .208        & .212        & .270       & .253       & .427     & .389   \\
KeyBERT                        & .150        & .209\ssa    & .217       & .183       & .405     & .386\ssa & .192        & .199        & .274       & .246       & .449     & .397   \\
KeyBERT + RM3                    & .151        & .220\ssa    & .203       & .175       & .385     & .371\ssa & .218        & .228\ssa    & .262       & .254       & .401     & .384   \\ \hline
QGMT                           & .170        & .238\ssa        & .236       & .197       & {.470}     & .393     & .232\ssa        & .265\ssa        & .346       & .273       & .547     & .471   \\
QGMT + RM3                       & .181        & \textbf{.257}\ssa        & .265       & .221       & .414     & .409     & \textbf{.262}\ssa        & .301\ssa        & \textbf{.368}  & \textbf{.323}       & .534     & \textbf{.498 }  \\ \hline
QGGT (Initial)               & .151        & .227        & .211       & .167       & .394     & .326\ssa & .234        & .269\ssa    & .340       & .303       & .526     & .452   \\
QGGT (Initial) + RM3                       & .166        & .244\ssa    & .217       & .188       & .405     & .342\ssa & .260\ssa    & .297\ssa    & .364       & .318       & .499     & .477   \\
QGGT (Initial \& Refined) & .132        & .220        & .204       & .165       & .403     & .307\ssa & .221        & .281\ssa    & .314       & .278       & \textbf{.566}     & .417   \\
QGGT (Initial \& Refined)  + RM3                       & .146        & .245\ssa    & .229       & .191       & .444     & .348\ssa & .243\ssa    & \textbf{.307\ssa}    & .344       & .296       & .506     & .441  
\end{tabular}%
}
\caption{Retrieval performance across the two benchmark collections, by the prompts related to query generation for clinical trials retrieval.}
\label{tab:query_gen_resutls}
\end{table}

The remaining rows of Table~\ref{tab:query_gen_resutls} show the retrieval performance achieved by the usage of the ChatGPT generated queries with BM25 and the standard PyTerrier query processing steps.
We remind that the acronym \textit{QGMT} stands for \textbf{Q}uery \textbf{G}eneration, \textbf{M}edical Role \& \textbf{T}ask Description.
In this approach, thoroughly described in Section~\ref{sec:QGMT}, the prompt provides contextual information to ChatGPT and instructs it to generate a single keyword-based query.
The acronym \textit{QGGT} refers to \textbf{Q}uery \textbf{G}eneration, \textbf{G}eneric Role \& \textbf{T}ask Description prompts presented in Section~\ref{sec:QGGT}.
Here, a single prompt is employed to generate two distinct keyword-based queries namely, \textit{Initial} and \textit{Refined}.
We run a total of four experiments; two exploit the \textit{Initial} query, and two exploit the concatenation of the \textit{Initial} and the \textit{Refined} queries.
In our experiments, we found that using solely the \textit{Refined} query, which mostly consists of novel terms (not included in the original query), lead to great decreases in performance across all measures and collections.
Specifically, P@10 for the TREC 2021 collection was .195 and .212 when RM3 is used for query expansion; similarly for the TREC 2022 collection.
This finding suggests that the new terms added to the \textit{Refined} query have probably a broader semantic meaning and lead to topic drift.

The QGMT experiment outperformed all of the QGGT related experiments in the TREC 2021 collection and the QGMT + RM3 outperformed all of the QGGT related experiments in both collections. 
Also, QGMT + RM3 shows statistically significant improvements in Bpref for both TREC 2021 and TREC 2022 collections.
In general, this finding suggests that exploiting ChatGPT's variable that assigns a particular role to the AI system leads to different responses.
This finding is inline with related studies~\cite{DBLP:journals/corr/abs-2304-05335}.
To further support the validity of the above finding, we run the QGMT experiment by setting the system's role to generic, i.e. ``You are a helpful assistant.''
The obtained results are presented in Table~\ref{tab:setting_systemrole}. It can be seen that when the system's role is generic, the retrieval performance decreases.
By observing the generated queries, we noticed that the model's behavior has also changed; specifically, with the generic role, the first extracted terms for the majority of the queries are the patient's age and gender. When the role is the domain-specific, the first extracted terms are related to the patient's medical problem.
\begin{table}[!htbp]
\resizebox{1\textwidth}{!}{%
\begin{tabular}{c|cccccc|cccccc}
                              & \multicolumn{6}{c|}{TREC 2021}                                                 & \multicolumn{6}{c}{TREC 2022}                                                    \\ \hline
                              & Rprec & Bpref & P@10 & P@25 & MRR & nDCG@10 & Rprec & Bpref & P@10 & P@25 & MRR & nDCG@10   \\ \hline
Employed Prompt (see Section~\ref{sec:QGMT}) & .170        & .238        & .236       & .197       & {.470}     & .393     & .232        & .265        & .346       & .273       & .547     & .471   \\
(System's role: as in Table~\ref{tab:query_generation}) &      &        &        &       &      &    &         &       &      &     & & \\ \hline
Employed Prompt (see Section~\ref{sec:QGMT})    & .160        & .195        & .215       & .189      & .454     & .443  & .216        & .211    & .324       & .281       & .518     & .468     \\ 
(System's role: ``You are a helpful assistant'') &      &        &        &       &      &    &         &       &      &     & & \\
\end{tabular}%
}
\caption{Retrieval performance across the two benchmark collections, using the prompt presented in Section~\ref{sec:QGMT} with different system's role.}
\label{tab:setting_systemrole}
\end{table}
To conclude, our findings suggest that instructing ChatGPT to directly generate queries for clinical trials retrieval has some potential, however the generated queries lead to inconsistent retrieval results across the collections. 
Also, by experimenting with the assigned role to the AI system, we observed that it is better if the role is related to the domain.
\subsection{ChatGPT for information extraction}
\label{sec:res_ie}
This section presents the results obtained by the prompts presented in Sections~\ref{sec:general_extraction} and~\ref{sec:domspec_extraction}.
Both of these approaches guide ChatGPT to perform information extraction from a given text, and their main differences rely on the number of system interactions and the level of the provided contextual information (i.e. related to the task and the system's role).
\begin{table}[!htbp]
\setlength\tabcolsep{0.5em}
\resizebox{1\textwidth}{!}{%
\begin{tabular}{c|cccccc|cccccc}
                              & \multicolumn{6}{c|}{TREC 2021}                                                 & \multicolumn{6}{c}{TREC 2022}                                                    \\ \hline
                              & Rprec & Bpref & P@10 & P@25 & MRR & nDCG@10 & Rprec & Bpref & P@10 & P@25 & MRR & nDCG@10   \\ \hline
BM25                          & .162        & .184        & .264       & .211       & .471     & .469   & .180        & .172        & .272       & .235       & .507     & .394     \\
BM25 + RM3                      & .184        & .241\ssa    & .285       & .231       & .463     & .484   & .208        & .212        & .270       & .253       & .427     & .389     \\
KeyBERT                       & .150        & .209        & .217       & .183       & .405     & .386   & .192        & .199        & .274       & .246       & .449     & .397     \\
KeyBERT + RM3                   & .151        & .220        & .203       & .175       & .385     & .371   & .218        & .228        & .262       & .254       & .401     & .384     \\ \hline
IEG (Extracted)               & .172        & .214\ssa    & .260       & .219       & .486     & .442   & .234\ssa    & .229\ssa    & .360\ssa   & .300\ssa   & .561   & .495\ssa \\
IEG (Extracted) + RM3                       & .196\ssa    & .258\ssa    & .299       & .243       & .498     & .463   & .252\ssa    & .278\ssa    & .360       & {.322\ssa}   & \textbf{.618 }  & .495\ssa \\
IEG (Extracted \& Expanded) & .163        & .224\ssa    & .252       & .205       & .500     & .424   & .211        & .242\ssa    & .350       & .294       & .570     & .476     \\
IEG (Ext. \& Exp.) + RM3                       & .165        & .258\ssa    & .248       & .203       & .470     & .414   & .240        & .282\ssa    & .346       & .309\ssa   & .538     & .472     \\ \hline
IEMT                          & .195\ssa    & .250\ssa    & .273       & .240       & .502     & .470   & .250\ssa    & .270\ssa    & .358\ssa   & .308\ssa   & .609   & .505\ssa \\
IEMT + RM3                      & \textbf{.212\ssa}    & \textbf{.275\ssa}    & \textbf{.323\ssa }  & \textbf{.261\ssa}   & \textbf{.541 }    & \textbf{.512}   & \textbf{.276\ssa}    & \textbf{.298\ssa}    & \textbf{.372\ssa}   & \textbf{.338\ssa}   & .576     & \textbf{.517\ssa}
\end{tabular}%
}
\caption{Retrieval performance across the two benchmark collections, using prompts that guide ChatGPT to extract keywords.}
\label{tab:information_extraction}
\end{table}
We remind that the first four rows of Table~\ref{tab:information_extraction} are identical among all result tables as they concern our baselines.
The \textit{IEG} abbreviation stands for \textbf{I}nformation \textbf{E}xtraction, \textbf{G}eneric \textbf{R}ole and refers to the prompt presented in Section~\ref{sec:general_extraction}; the \textit{IEMT} (\textbf{I}nformation \textbf{E}xtraction, \textbf{M}edical Role \& \textbf{T}ask Description) refers to the prompt presented in Section~\ref{sec:domspec_extraction}.

Inline with the previous findings, providing the AI system with clear domain and task information leads to better responses and as a result to better retrieval performance. In detail, the IEMT experiment outperforms both the IEG (Extracted) and the IEG (Extracted \& Expanded) across all measures and both collections.
In addition, IEMT performs equally or better to the BM25 + RM3 baseline in TREC 2021 collection and outperforms it in TREC 2022. From these results, the observed increase in Bpref is statistical significant in both collections.
With the incorporation of the RM3 model for query expansion, we observe that the performance of IEMT (experiment IEMT + RM3) shows a statistical significant increase over the BM25 baseline for the majority of the presented measures and across both of the employed collections.

Focusing on the IEG (Extracted) and the IEG (Extracted \& Expanded) experiments, one can observe that the later decreases the retrieval performance across most of the presented measures.
This finding further supports our observation that when ChatGPT is instructed to expand or refine its response, it tends to add terms with a broader semantic meaning leading to topical drift and performance decrease.
\subsection{ChatGPT for information extraction, meaning disambiguation and query formulation}
\label{sec:res_ecir}
By leveraging the prompt presented in Section~\ref{sec:ie_meaning} we extracted and disambiguate the meaning of various patient-related information from a clinical note.
Then, based on a previous study~\cite{DBLP:conf/ecir/PeikosAPV23}, we synthesized queries by combining the extracted information in various possible combinations.
To give an overview, we estimated the retrieval performance achieved by combining identified diseases and medical problems with symptoms, or combining diagnosis, problems and diseases with treatments, and many other combinations.
In this section, we present the results obtained by the query that achieved the higher Rprec in both collections (experiment IEMDMT).
We remind that \textit{IEMDMT} stands for \textbf{I}nformation \textbf{E}xtraction \& Entity \textbf{M}eaning \textbf{D}isambiguation, \textbf{M}edical Role \& \textbf{T}ask Description.
\begin{table}[!htbp]
\setlength\tabcolsep{0.5em}
\resizebox{1\linewidth}{!}{%
\begin{tabular}{c|cccccc|cccccc}
            & \multicolumn{6}{c|}{TREC 2021}                                                   & \multicolumn{6}{c}{TREC 2022}                                                  \\ \hline
            & Rprec & Bpref & P@10 & P@25 & MRR & nDCG@10   & Rprec & Bpref & P@10 & P@25 & MRR & nDCG@10 \\ \hline
BM25        & .162        & .184        & .264       & .211       & .471     & .469     & .180        & .172        & .272       & .235       & .507     & .394   \\
BM25 + RM3    & .184        & .241\ssa    & \textbf{.285}    & \textbf{.231}       & .463     & \textbf{.484}     & .208        & .212        & .270       & .253       & .427     & .389   \\
KeyBERT     & .150        & .209        & .217       & .183       & .405     & .386     & .192        & .199        & .274       & .246       & .449     & .397   \\
KeyBERT + RM3 & .151        & .220        & .203       & .175       & .385     & .371     & .218        & .228        & .262       & .254       & .401     & .384   \\ \hline
IEMDMT      & .174        & .238\ssa    & .237       & .211       & \textbf{.485}     & .397\ssa & .247\ssa    & .260\ssa    & \textbf{.348\ssa}   & .285       & .528     & .458   \\
IEMDMT + RM3  & \textbf{.185}        & \textbf{.268\ssa}    & .264       & .222       & \textbf{.485}     & .428     & \textbf{.258\ssa}    & \textbf{.285\ssa}    & .328       & \textbf{.311\ssa }  & \textbf{.560}     & \textbf{.461}  
\end{tabular}%
}
\caption{Retrieval performance across the two benchmark collections, by a query that is formulated by combing the extracted 'diagnosis', 'medical\_problem', 'diseases', 'drug', 'symptoms', 'treatments', 'medications', 'lab\_examinations', identified by ChatGPT and expanded by suggested 'MeSH\_terms'.}
\label{tab:ecir_insp}
\end{table}
The query contains the unique terms obtained by concatenating the extracted medical diagnosis, problem, diseases, drugs, symptoms, treatments, medications, examinations, along with the generated MeSH terms. 
Although Wang et al. noticed that ChatGPT suggests non-existing or poor quality MeSH Terms~\cite{DBLP:journals/corr/abs-2302-03495}, our experiments suggest that including the suggested MeSH terms in the query leads to small percentage improvements in performance (around 6\% in TREC 2021 and 9\% for TREC 2022). 
In contrast, resolving the abbreviations hurt retrieval effectiveness.
However, we leave a more detailed analysis related to the quality of the suggested MeSH terms and the accuracy of the resolved abbreviations for future work.
Including information related to a patient's past medical history, family history, lab results, drug dosages, and lifestyle factors in the synthesized query, lead to decreases in performance.

Based on previous observations reported in Section~\ref{sec:ie_meaning}, we suspect that the underline reasons for the observed performance decreases might be related to the following reasons.
First, decreases might be due to the model's responses when a clinical note did not contain information related to the past medical history, family history, e.g. \textit{``not applicable.''}
In fact, ChatGPT did not identify information about a patient's family or medical history for several clinical notes used in our study.
Another reason might be related to the inclusion of negated answers, that have been observed during the extraction of a patient's lifestyle factors.
To conclude, the reported retrieval performance using this prompt, might have been underestimated due to the format of the model's responses.
Therefore, further research should improve the IE extraction process, by perhaps exploiting another prompting technique or extracting the information with more interactions with the system.
\subsection{Negation removal and ChatGPT for information extraction}
\label{sec:res_nr_ie}
The results presented in this section correspond to the hybrid approach, presented in Section~\ref{sec:bert_chat_section}, that exploits a pre-trained language model for negation identification and ChatGPT for information extraction.
We remind that the presented experiment \textit{NRIEMT}, has the same prompt for IE with ChatGPT with the IEMT experiment presented in Section~\ref{sec:res_ie}. This prompt instructs the model to extract information by providing all contextual information related to the task and the system's role.
Therefore, Table~\ref{tab:non_neg} repeats the retrieval performance achieved by IEMT, for comparison purposes.
\begin{table}[!htbp]
\setlength\tabcolsep{0.5em}
\resizebox{1\textwidth}{!}{%
\begin{tabular}{c|cccccc|cccccc}
            & \multicolumn{6}{c|}{TREC 2021}                                                 & \multicolumn{6}{c}{TREC 2022}                                                    \\ \hline
            & Rprec & Bpref & P@10 & P@25 & MRR & nDCG@10 & Rprec & Bpref & P@10 & P@25 & MRR & nDCG@10   \\ \hline
BM25        & .162        & .184        & .264       & .211       & .471     & .469   & .180        & .172        & .272       & .235       & .507     & .394     \\
BM25 + RM3    & .184        & .241\ssa    & {.285}       & .231       & .463     & {.484}   & .208        & .212        & .270       & .253       & .427     & .389     \\
KeyBERT     & .150        & .209        & .217       & .183       & .405     & .386   & .192        & .199        & .274       & .246       & .449     & .397     \\
KeyBERT + RM3 & .151        & .220        & .203       & .175       & .385     & .371   & .218        & .228        & .262       & .254       & .401     & .384     \\ 
\hline
IEMT  & .195\ssa    & .250\ssa    & .273       & .240       & .502     & .470   & .250\ssa    & .270\ssa    & .358\ssa   & .308\ssa   & .609   & .505\ssa \\
IEMT + RM3                      & \textbf{.212\ssa}    & \textbf{.275\ssa}    & \textbf{.323\ssa }  & \textbf{.261\ssa}   & \textbf{.541 }    & \textbf{.512}   & \textbf{.276\ssa}    & \textbf{.298\ssa}    & {.372\ssa}   & \textbf{.338\ssa}   & .576     & \textbf{.517\ssa}\\
\hline
NRIEMT      & .185        & .229\ssa    & .260       & .221       & {.508 }    & .449   & .247\ssa    & .249\ssa    & \textbf{.374\ssa}  & .315\ssa   & .598     & .496\ssa \\
NRIEMT + RM3  & {.191}        & {.255\ssa}    & .277       & {.241}       & .491     & .466   & {.272\ssa }   & {.285\ssa}    & .372\ssa   & \textbf{.338\ssa}   & \textbf{.623}     & {.509\ssa}
\end{tabular}%
}
\caption{Retrieval performance across the two benchmark collections, by the prompts related to keyword extraction applied after negated content has been removed from the original queries. For comparison purposes we present also the IEMT prompt setting.}
\label{tab:non_neg}
\end{table}

The NRIEMT experiment aims to investigate whether removing negated information prior to performing IE for a clinical note leads to better retrieval performance based on the premise that the generated queries will be focused on patient existing condition.
An effectiveness indicator would have been an increase in all measures, but specifically in MRR, P@10 and P@25. 
However, this is not the case in our experiments.
Firstly, both NRIEMT and NRIEMT + RM3 experiments, show statistically significant improvements over the BM25 baseline in both collections, suggesting that this prompting method is stable.
By comparing the IEMT and NRIEMT experiments, one can observe that the results are not consistent across collections.
For TREC 2021, precision drops, while in TREC 2022 we have small and not statistically significant improvements.
A plausible reason might be the characteristics of the queries in each collection, i.e. for TREC 2022 there might be more negated medical entities and their removal might positively impact retrieval performance for those queries.
Another reason, might be that ChatGPT generated slightly different responses (do not extract the same terms as IEMT), especially for queries from which various information have been removed. 
In any case, the presented findings encourage further exploration towards this research direction, i.e. how one can exploit LLMs for negation handling in clinical text.  
\section{Summary of Findings and Discussion}
\label{sec:find_disc}
In this section we summarize the main contributions and findings of our investigation.
Firstly, to summarize the findings of our work, we rank the conducted experiments based on their achieved Rprec measure (see Table~\ref{tab:exp_ranking}).
We select Rprec as it has been empirically found that it is highly correlated with MAP~\cite{DBLP:books/daglib/0021593} and therefore provides a better understanding of the quality of the obtained document ranking.
Then, we compare our best performing experiment with an information extraction approach proposed in the literature~\cite{DBLP:conf/ecir/PeikosAPV23}, and the TREC 2021 and 2022 SoA approaches. 
Finally, we conduct additional experiments to investigate if ChatGPT generated queries can reach the retrieval performance achieved when human-generated queries are used for clinical trials retrieval.

By ranking our experiments, one notices that for TREC 2022, all of the conducted experiments outperform both of the selected baselines (BM25 and KeyBERT).
For TREC 2021 one can notice that only the information extraction prompts lead to better Rprec values.
Across collections, the prompts that instruct ChatGPT to extract information from the clinical notes (those whose abbreviation starts with ``IE'' and ``NRIE,'') lead to greater Rprec values than those that instruct it to generate queries (abbr. starts with ``QG'').
The only exception is the IEG (Extracted \& Expanded) experiment, which in TREC 2021 underperforms compared to BM25 and in  TREC 2022 it has the lower Rprec value among our experiments.
Thus, ChatGPT can be used to process unstructured clinical notes, but it is preferred to be instructed to extract patient-related information rather than generate queries for clinical trials retrieval. 
\begin{table}[!htbp]
\centering
\setlength\tabcolsep{0.5em}
\resizebox{.7\textwidth}{!}{%
\begin{tabular}{c|c}
Rprec -   TREC 2021                 & Rprec - TREC 2022                 \\ \hline
QGGT (Initial \& Refined)   [.132]   & \textbf{BM25 [.180]}              \\
KeyBERT [.150]                      & KeyBERT [.192]                    \\
QGGT (Initial) [.151]               & IEG (Extracted \& Expanded) [.211] \\
IEG (Extracted \& Expanded)   [.162] & QGGT (Initial \&  Refined) [.221]   \\
\textbf{BM25 [.162]}                & QGMT [.232]                       \\
QGMT [.170]                         & QGGT (Initial) [.234]             \\
IEG(Extracted) [.172]               & IEG(Extracted) [.234]             \\
IEMDMT [.174]                       & IEMDMT [.247]                     \\
NRIEMT [.185]                       & NRIEMT [.247]                     \\
IEMT [.195]                         & IEMT [.250]          
\end{tabular}%
}
\caption{Ranking of experiments based on their achieved Rprec across the two collections. See Table~\ref{tab:experiments} for more information about the used abbreviations. The value in brackets is the achieved Rprec.}
\label{tab:exp_ranking}
\end{table}

Another finding is highlighted by the Rprec achieved by the IEG (Extracted \& Expanded) and the QGGT (Initial \& Refined) experiments.
As it can be seen in the table, both of these experiments have poor performances in both collections.
We remind that in these prompts, ChatGPT has been instructed to ``Expand the extracted list of keywords with 20 additional related keywords'' (IEG) and ``Refine the query based on further details, such as synonyms or related medical concepts.'' (QGGT).
Through the presented examples we noticed that the selected expansion terms are related to more generic medical concepts than those mentioned in the original clinical note.
Therefore, based on the experiments conducted in this study, it is noted that query expansion with ChatGPT leads to poor performance.
Future studies should further investigate towards this direction aiming to investigate whether ChatGPT can be instructed to generated more specific medical concepts.

As mentioned in previous sections, the IEMDMT experiment has been inspired by a previous study~\cite{DBLP:conf/ecir/PeikosAPV23}, in which the authors extracted various patient-related information from a clinical note using previous state-of-the-art methods and used these pieces of information to synthesize queries.
In their work, the authors conduct their experiments on the TREC 2021 collection and report Bpref, P@5, and P@25.
Therefore, Table~\ref{tab:ecir_here} compares their reported results to those achieved by the IEMDMT and the IEMT experiments.
\begin{table}[!htbp]
\centering
\resizebox{.4\textwidth}{!}{%
\setlength\tabcolsep{0.5em}
\begin{tabular}{c|ccc}
       & \multicolumn{3}{c}{TREC 2021} \\ \hline
       & Bpref    & P@5      & P@25    \\ \hline
\begin{tabular}[c]{@{}c@{}}Previous Study~\cite{DBLP:conf/ecir/PeikosAPV23}\\       \end{tabular} & .214 & .331 & .227 \\ \hline
IEMDMT & .238    & .269    & .211   \\
IEMT   & .250    & .331    & .240 
\end{tabular}%
}
\caption{Comparison to the previous study presented in~\cite{DBLP:conf/ecir/PeikosAPV23}.}
\label{tab:ecir_here}
\end{table}
As it can be seen in the table, the IEMDMT prompt improves the retrieval performance over the previous study only in terms of Bpref.
This finding suggests that ChatGPT, with the prompt used in this study, does not extracts patient-related information as accurately as the domain-specific state-of-the-art methods employed in~\cite{DBLP:conf/ecir/PeikosAPV23}.
However, in previous sections we have identified various reasons that might have underestimated the performance achieved by IEMDMT and point out various research directions that might improve it.
Nonetheless, the best performing prompt, i.e. IEMT, improves or has the same performance as the one reported in~\cite{DBLP:conf/ecir/PeikosAPV23}.
To conclude, it seems that the employed prompt (IEMDMT) is not sufficient to capture all the semantic information that is present in a clinical note, as good as other methods employed in the literature.

Regarding the NRIEMT experiment, it did not meet our expectation to improve the performance of the IEMT experiment by removing negated medical entities.
In addition, our attempts to remove the negated content of a clinical note using ChatGPT, highlighted some warning behavior.
Specifically, as we have briefly mentioned in Section~\ref{sec:bert_chat_section}, we have observed that ChatGPT altered the meaning of the original text, rather than removed the negated medical entities.
In the studied task, this behavior might has a small impact, as it will only lead to a performance decrease during clinical trials retrieval.
However, as the employment of LLMs as assistants to conduct medical bureaucratic tasks is supported by various organizations, this behavior might cause greater problems in other applications.

\subsection{Comparison with the SoA approaches}
This sections compares the retrieval performance achieved by the top performing experiments conducted in our research study with the SoA approaches in TREC 2021 and TREC 2022 (see Table~\ref{tab:soa_comp}).
Unfortunately, our understanding around the TREC 2022 SoA approach is limited as the participating team did not provide a detailed description of their experimental design.

The SoA approach of TREC 2021, conducted by the same team, is described in~\cite{10.1145/3477495.3531853} and it has been briefly analyzed in Section~\ref{sec:related_work}.
The approach is a multi-stage neural ranking approach.
Here, we further analyze the approach by focusing on the initial retrieval phase.
Specifically, given a clinical note, the authors leverage a neural query synthesis (NQS) method (i.e. a zero-shot document expansion model) to generate forty sentence-long queries.
Each of these queries is used independently, in addition to the original clinical note, as input in a retrieval pipeline that exploits the BM25 and RM3 models, and the obtained results are fused.
The aforementioned process is their first-stage retrieval method, whose results are presented in the second row of Table~\ref{tab:soa_comp} (``Neural Query Synthesis'').
\begin{table}[!htbp]
\centering
\resizebox{\textwidth}{!}{%
\setlength\tabcolsep{0.5em}
\begin{tabular}{cccc|cccc}
\multicolumn{4}{c|}{TREC   2021}                    & \multicolumn{4}{c}{TREC 2022}          \\ \hline
Run                     & nDCG@10 & P@10   & MRR     & Run          & nDCG@10 & P@10  & MRR    \\ \hline
TREC's Median                  & .304   & .1613 & .2942 & TREC's Median       & .392   & .258 & .411 \\ \hline
Neural Query Synthesis~\cite{10.1145/3477495.3531853}  & .473   & .276  & .434  & \textit{frocchio} run & .463   & .324 & .537 \\
monoT5\textit{'CT}~\cite{10.1145/3477495.3531853} & \textbf{.712} & \textbf{.593} & \textbf{.816} & \textit{frocchio\_monot5\_e} run & \textbf{.613} & \textbf{.508} & \textbf{.726} \\ \hline
IEMT+RM3                & .512   & .323  & .541  & NRIEMT + RM3 & .509   & .372 & .623
\end{tabular}%
}
\caption{Comparison between our best performing experiments with the state-of-the-art approaches of TREC 2021 and TREC 2022.}
\label{tab:soa_comp}
\end{table}
Then, a neural re-ranker based on the monoT5 model, fine-tuned for the task of clinical trials retrieval is employed to create the final ranking.
The achieved performance is presented in the third row of Table~\ref{tab:soa_comp} (``Multi-Stage monoT5\textit{'CT}'').
As it can be seen, the IEMT+RM3 experiment leads to better retrieval effectiveness compared to the NQS (first-stage retrieval) approach.
We further note that even the IEMT experiment reaches similar performance for nDCG@10 and P@10 compared to the proposed NQS approach. Also, IEMT reaches higher MRR (.502) than NQS.
In addition, the IEMT and IEMT+RM3 experiments are single and two-stage retrieval approaches, respectively.
In contrast, the NQS approach necessitates forty initial retrieval runs, involving the use of generated query variations. Consequently, our approaches exhibit lower complexity when compared to NQS.

The SoA approach in the TREC 2022 Clinical Trials track, also employs the Mono-T5 model and it has been proposed by the same research team. Due to the lack of provided experimental details, it is not feasible to further comment on the details of this approach. 
However, based on the submitted experiments, we can see that our best performing experiment in the TREC 2022 collection outperforms the \textit{frocchio} run, that was probably the employed first-stage retrieval (based on the run name).
Nonetheless, also the IEMT experiment outperforms the \textit{frocchio} run in all of the presented retrieval measures.

To conclude, based on the comparisons presented in Table~\ref{tab:soa_comp}, the experiments conducted in this work outperformed the first-stage retrieval approaches exploited in both the TREC 2021 and 2022 study.
However, our experiments did not yield any performance improvements compared to the SoA performance achieved by the monoT5 based neural re-ranker.
Nonetheless, when combined with our retrieval approach, these neural re-ranking models may enhance their retrieval performance.
\subsection{Comparison with human-generated queries}
\label{sec:comp_human}
The objective of this section is to investigate if the retrieval performance achieved with ChatGPT-generated queries can match that of human-generated queries in the context of clinical trials retrieval.
To achieve that we exploit the retrieval collection introduced by Koopman and Zuccon~\cite{DBLP:conf/sigir/KoopmanZ16}.
The provided collection bears resemblance to the ones used in the TREC 2021 and TREC 2022 clinical trials tracks.
In line with these tracks, a total of 60 queries are provided in this collection, mimicking patient cases in the form of admission statements in EHRs.
Additionally, the document collection comprises clinical trials extracted from ClinicalTrials.gov.
Finally, each document (i.e. clinical trial) has been assessed in a three-graded scale, similarly to the TREC tracks, where a label of 2 means that the patient is eligible to participate in the given trial.

In their study, Koopman and Zuccon~\cite{DBLP:conf/sigir/KoopmanZ16} asked four medical assessors to provide, for each patient case, several ad-hoc
keyword queries that they would issue to a search engine
to find clinical trials.
As reported, the final collection contains a total of 489 unique queries, with 8.2 keyword queries per patient case created by the four medical assessors.
Hereafter we refer to the medical assessors as assessor A, B, C, and D.

To investigate whether the ChatGPT-generated queries can match that of human-generated queries we conduct the following experiments.
First, for each patient case in the collection, we concatenate the $n$ ad-hoc keyword queries created by each individual assessor into a single ad-hoc keyword query which contains all the unique keywords.
For example, assuming that the assessor A has created two distinct ad-hoc keyword queries 
\textit{q1} = \textit{keyword1}, \textit{keyword2}, \textit{keyword3} and
\textit{q2} = \textit{keyword2}, \textit{keyword4} for a specific patient case, the final query that is used for retrieval will be \textit{qf} = \textit{keyword1}, \textit{keyword2}, \textit{keyword3}, \textit{keyword4}.
The created \textit{qf}, simulates a scenario in which an expert user is presented with a patient's case and is asked to think of all possible clinical trials this patient can participate.
Following this process for each assessor we have created four ad-hoc queries for each patient case.
Using the distinct created queries \textit{qfA}, \textit{qfB}, \textit{qfC}, and \textit{qfD} as inputs, we run four retrieval experiments based on the same experimental set-up presented in Section~\ref{sec:exp_setup}.
However, medical assessors A, B and C have not provided ad-hoc keyword queries for 1, 3, and 1 patient cases respectively.
In the conducted experiments, these patient cases have been removed from the evaluation.

In addition, we concatenate \textit{qfA}, \textit{qfB}, \textit{qfC}, and \textit{qfD} into a single query that contains their unique keywords, i.e. \textit{qfall} and used this as input in another retrieval experiment.
We assume that the \textit{qfall} query accumulates the knowledge of various medical experts in a single representation. 
Lastly, to generate queries with ChatGPT we employed the IEMT approach presented in Section~\ref{sec:domspec_extraction}.

To compare the retrieval effectiveness achieved by the human-generated queries with that of ChatGPT generated we report the obtained P@10 and the Bpref measure.
For both measures we assume only the \textit{eligible} clinical trials as relevant, i.e. trials assessed as \textit{excludes} are considered as \textit{not relevant}.
As the authors report, in this benchmark collection, the number of assessed documents is limited and therefore the evaluation might be less reliable for new systems.
To overcome this issue, we evaluate the retrieval performance by employing the condensed measures approach proposed in~\cite{DBLP:conf/sigir/Sakai07} as a way to deal with retrieved but unjudged documents. 
The obtained results are reported in Table~\ref{tab:chat_vs_human}.

\begin{table}[!htpb]
\centering
\resizebox{.4\textwidth}{!}{%
\setlength\tabcolsep{0.5em}
\begin{tabular}{c|cc}
Experiment     & Bpref & P@10 (Condensed) \\ \hline
\textit{qfA}            & .117 & .140            \\
\textit{qfB}            & .093 & .120            \\
\textit{qfC}            & .059 & .100            \\
\textit{qfD}            & .116 & .138            \\
\textit{qfall}          & .090 & .110            \\ \hline
IEMT           & .107 & .131           
\end{tabular}%
}
\caption{Retrieval performance achieved by the human-generated and ChatGPT-generated queries.}
\label{tab:chat_vs_human}
\end{table}

Based on the table, IEMT leads to better retrieval performance compared to the accumulated queries of assessors B, C and all combined (i.e. \textit{qfall}).
However, it fails to improve retrieval performance against assessors A and D.
It is reported in~\cite{DBLP:conf/sigir/KoopmanZ16} that assessors B and C, i.e. those whose queries unperformed compared to ChatGPT's, had a different behavior than assessor A and D.
Specifically, assessors B and C created many small queries, while assessors A and D created fewer but longer ones (i.e. contained more keywords).
These findings suggest that ChatGPT might be better for information extraction for this tasks compared to medical experts, under certain circumstances.
Nonetheless, more experiments should be conducted to further investigate whether ChatGPT, or another LLM, can generate better queries for clinical trials retrieval than medical experts.

\section{Limitations and Potentials of ChatGPT}
\label{sec:limitations}
In this section, we report the limitations encountered in our experiments. Additionally, we examine the potential positive and negative impacts of utilizing the proposed approach for automating the task of clinical trials retrieval.
\subsection{Usage and experimental limitations of ChatGPT in this study}
This study uses ChatGPT (specifically the gpt-3.5-turbo model) by leveraging OpenAI's API. Our experiments were conducted over two weeks starting after March 16th 2023. It is important to note that ChatGPT is treated as a black box in this study, as details regarding its architecture and the specific training data used are undisclosed. Moreover, we remind that each query is processed in a separate system interaction via the API, i.e. through a new conversation chat.
According to the documentation, ChatGPT has been trained using data available until September 2021. However, it remains uncertain whether the model has specifically been trained on the TREC 2021 topics used in this study, which were published in May 2021, or on similar topics such as those discussed in a previous work~\cite{DBLP:conf/sigir/KoopmanZ16}. In addition, the model has probably not been exposed to or trained on the TREC 2022 topics, as they were published on June 2, 2022.
The behaviour of the ChatGPT model is non-deterministic, meaning that it can produce different responses for the same input. To mitigate this, we set the \textit{temperature} parameter to zero in our experiments, intending to make the generated responses as deterministic as possible. However, as mentioned in the documentation, a slight amount of variability may still be present.
Additionally, it is essential to note that all of our experiments were performed using a single user account. Therefore, whether the provider retains user-related information from previous system conversations and uses it to personalize the model's responses is uncertain.
In order to examine the response variability of the model, we conducted a repeat of the IEMT experiment on April 24 2023, which was one month after the initial experiment. It is important to note that the same user account was utilized for both experiments. The retrieval performance achieved in the two experiments was found to be identical. This result suggests that the IEMT prompting approach exhibits robustness, at least when the same user account is employed. However, future work should further study this aspect to understand ChatGPT's response variability better. 
\subsection{Potentials of using ChatGPT for clinical trials retrieval}
In this research, ChatGPT is used as an intermediary tool to facilitate information extraction from unstructured clinical notes, a task that traditionally demands substantial human involvement. In this section, we highlight the benefits of employing ChatGPT, as described in our study, to automate or semi-automate this essential stage of the clinical enrollment workflow.
The advanced language processing capabilities of ChatGPT enable efficient extraction of the desired information from clinical notes, minimizing the required human effort and mitigating the risk of human errors.
Concurrently, our findings indicate that queries generated by ChatGPT result in improved retrieval performance compared to existing approaches in the literature, and, occasionally, those generated by human experts.
The capabilities of ChatGPT indicate that it can replace multiple domain-specific language models that are fine-tuned for extracting specific pieces of information. Consequently, ChatGPT will simplify the operation and maintenance of an information extraction system designed to process unstructured clinical notes.
Concerning the issue of hallucination generation, the proposed approach mitigates this concern by restricting the model's focus to information extraction rather than attempting to answer medical queries or make decisions about a patient's medical situation.
In conclusion, ChatGPT has the potential to automate the information extraction process fully. However, it would be more suitable to incorporate human oversight and validation as an additional step to ensure the extracted information's accuracy, completeness, and reliability.
\subsection{Risks of Using ChatGPT for clinical trials retrieval}
\label{sec:risks}
One of the main limitations of ChatGPT and other generative language models is their non-deterministic behaviour.
In this study, we acknowledged this limitation and took significant steps to address it, although slight variations in model responses may still occur. In addition to non-determinism, other limitations should be considered in medical information extraction and retrieval, such as their lack of explainability and potential concerns related to data privacy.
Given the nature of the information extraction task in this study, where the focus is on extracting simple information rather than making decisions about a patient's situation, obtaining explanations from the model is of lesser significance.
It is worth noting that in the proposed approach ChatGPT does not make decisions about a patient's situation and its outputs do not directly impact patient care.
However, these limitations should be carefully considered when the model is expected to answer questions about a patient, such as determining whether the patient's condition is chronic or acute.
The process of extracting information from clinical notes using ChatGPT raises valid concerns regarding data privacy and security, as it involves handling sensitive patient information that needs to be protected.
To mitigate this risk, it would be essential to employ de-identification techniques on the electronic health records prior to extracting information from them, for example by using neural models~\cite{ahmed2020identification} or even LLMs~\cite{DBLP:journals/corr/abs-2303-11032}.
These de-identification methods help anonymize patient data, reducing the likelihood of exposing sensitive information during the extraction process. Additionally, it is crucial to implement the proposed approach in a manner that complies with regulations that ensure the security and confidentiality of patient-related information, such as the Health Insurance Portability and Accountability Act (HIPAA)\footnote{\href{https://www.cdc.gov/phlp/publications/topic/hipaa.html}{Health Insurance Portability and Accountability Act of 1996 (HIPAA).}}.
By adhering to these measures, healthcare providers can protect patient privacy while leveraging ChatGPT for information extraction purposes.
\section{Conclusions and Directions for Future Research}
\label{sec:conclusions}
The primary objective of this study is to improve the initial stage of the clinical enrollment workflow by incorporating ChatGPT as an intermediate component within a retrieval pipeline. ChatGPT extracts information from unstructured clinical notes that detail a patient's medical situation, employing specially designed prompts. The generated responses from ChatGPT serve as queries to identify relevant clinical trials in which the patient may be eligible to participate. The proposed approach explicitly addresses several limitations commonly associated with LLMs, including the potential for generating hallucinations and response inconsistency. Our investigation assesses the model's capabilities in query generation, information extraction, and disambiguation of essential patient-related information found in unstructured clinical notes.
Based on the empirical evaluation conducted, we provide a summary of the key conclusions of our study in terms of the research questions presented in this study.

Analyzing the results presented in Table~\ref{tab:exp_ranking}, it becomes evident that within the utilized prompting strategies, the IEMT experiments exhibit noteworthy enhancements in retrieval performance for both benchmark collections.
These experiments involve assigning a domain-specific role to the model and providing it with a comprehensive task description.
The observed improvements are statistically significant, highlighting the effectiveness of this approach.
Furthermore, the IEMT and the IEMT+RM3 experiments yield better retrieval performance compared to the state-of-the-art approaches for query generation (refer to Table~\ref{tab:soa_comp}) as well as human-generated queries (refer to Table~\ref{tab:chat_vs_human}).
These findings strongly support the conclusion that the patient-related information extracted using ChatGPT is sufficient to enhance retrieval performance in the studied search task and it also outperforms the SoA approaches and, occasionally, even human-generated queries.
In consideration of the final research question pertaining to the potential advantages of incorporating ChatGPT into the clinical trial enrollment pipeline, we contend that the answer is positive, provided that the employed system effectively addresses the risks outlined in Section~\ref{sec:risks} and protects patients privacy.

Our future work will address several shortcomings identified in our empirical experiments.
To begin with, in our experiments we noticed that when ChatGPT is instructed to expand or refine its response (i.e. generated query or extracted information), it tends to add terms with a broader semantic meaning leading to topical drift and performance decrease.
We would like to investigate towards this direction aiming to instruct ChatGPT to generated terms with requested semantic meaning. 
In addition to that, we intent to perform a detailed analysis related to the quality of the suggested MeSH terms and the correctness of the resolved abbreviations.
As mentioned in Section~\ref{sec:bert_chat_section}, the attempt to instruct ChatGPT to identify and remove negated content from a clinical note resulted in unexpected outcomes that completely changed the semantic meaning of a clinical note. Therefore, this model limitation requires further investigation in future studies, as it can potentially negatively impact the interpretation of clinical information.
Similarly, more extensive experiments will be conducted to investigate the models response variability over time, across different user profiles and slightly modified prompts.
As discussed in Section~\ref{sec:ie_meaning}, ChatGPT has demonstrated the ability to disambiguate the meaning of different medical terms within a clinical note. Based on this finding, we intend to leverage ChatGPT, or another domain-specific language model, such as GatorTron, to automate further an additional stage of the clinical enrollment workflow, specifically the eligibility screening process.

In summary, our study indicates that the utilization of ChatGPT has the potential to enhance the process of clinical trial enrollment.
By prioritizing data privacy and ensuring patient safety, this technology can contribute to improving the quality of medical services while reducing the burden on healthcare professionals.
\section*{Acknowledgements}
This work was supported by the EU Horizon 2020 ITN/ETN on Domain Specific Systems for Information Extraction and Retrieval (H2020-EU.1.3.1., ID: 860721).

\bibliographystyle{splncs04}
\bibliography{references}

\newpage
\appendix
\section{Additional prompts for query generation}
\label{sec:appenda}
Within this appendix, we introduce two prompting approaches that were assessed using the TREC 2021 benchmark collection.
Unfortunately, these approaches did not yield any improvement in retrieval performance compared to the approaches presented in our study. Moreover, they exhibited undesired behavior by generating hallucinations.
\subsection{A few-shot approach}
The initial approach is a few-shot approach, which makes use of four randomly selected queries from the benchmark collection presented by Koopman and Zuccon~\cite{DBLP:conf/sigir/KoopmanZ16}.
In their study, medical assessors were asked to provide multiple ad-hoc keyword queries that they would employ to search for clinical trials related to a specific patient case.
For this few-shot prompting approach, we utilize the original patient case along with the human-generated queries, as outlined in Table~\ref{tab:query_gen_refine_few_shot}.
In this approach, we randomly select four queries from the collection introduced in~\cite{DBLP:conf/sigir/KoopmanZ16} which are remain unchanged in every interaction with the system, i.e. new conversation chat.
This approach resembles the QGMT experiment as the model is assigned a particular role and a description of the search task is provided (see Section~\ref{sec:QGMT}).
The response example provided in Table~\ref{tab:query_gen_refine_few_shot} is consistent across the vast majority of the TREC 2021 queries.
As observed, the bi-gram "clinical trial" is added at the end of the extracted patient-related information.
\begin{table}[!htbp]
\setlength\tabcolsep{0.5em}
\begin{tabular}{p{10cm}|p{6cm}}
\multicolumn{1}{c|}{\centering Prompt Text} &
\multicolumn{1}{c}{\centering Model's output (example)} 
\\ \hline
Your task is to retrieve clinical trials from a collection of clinical trials in the United States. 
To achieve that, you have access to a clinical medical note of a patient. 
Do not explain or elaborate. 
Here are some examples of patient cases, and their associated created information by an expert. 
\newline\newline
Clinical note 1: A 58-year-old African-American woman presents to the ER with episodic pressing/burning anterior chest pain that began two days earlier for the first time in her life. ... She currently takes no medications. Physical examination is normal. The EKG shows nonspecific changes.",
\newline\newline
Output 1: 
Firstline management of new onset ACS in African-American Women, obese, Obesity ACS, Obesity IHD, chest pain trial, hypertension ACS, .... angina nitrate trial, stable angina beta blocker trial, stable angina calcium channel blocker trial, stable angina nitrate trial
\newline\newline
Clinical note 2: An 8-year-old male presents in March to the ER with fever up to 39 C, dyspnea and cough for 2 days. He has just returned from a 5 day vacation in Colorado. .... A chest x-ray shows bilateral lung infiltrates."
\newline\newline
Output 2: First line treatment of Colorado Tick Fever in children, Analgesic and Antipyretic options for treatment of Colorado Tick Fever in children, ... Travel gastroenteritis, pnuemonia antibiotic trial,
Clinical Trial Fever Cough Short of Breathe Children
\newline\newline
Clinical note 3: 64-year-old obese female with diagnosis of diabetes mellitus and persistently elevated HbA1c. ... She has tried using topical lotions and creams but the lesion has increased in size and is now oozing."
\newline\newline
Output 3: type two diabetes trial, diabetes poor adherence trial, diabetes poor compliance trial, diabetic ulcer trial, diabetic ulcer clinical trial, ... Clinical Trial Diabetes Wound oozing not healing, Neurogenic ulcerations in non-compliant diabetes paitents
\newline\newline
Clinical note 4: A 28-year-old female with neck and shoulder pain and left hand and arm paresthesias three weeks after returning from a trip to California  ... increasing agitation and anxiety, malaise, difficultly swallowing and marked hydrophobia, and was immediately hospitalized.",
\newline\newline
Output 4: rabies, clinical Trial Parathesis arm stray animal tremors, Clinical Trial increasing paralysis swallowing, malaise spastic arm movements,... Prognostic factors for unidentified rabies virus, rabies trial
\newline\newline
Clinical note: {\textit{Clinical Note}}
&
``anaplastic astrocytoma clinical trial'',\newline ``spinal cord tumor clinical trial'', \newline``radiation therapy clinical trial'',\newline ``temozolomide clinical trial'',\newline ``CPT-11 clinical trial'',\newline ``Avastin clinical trial'', \newline``lower extremity weakness clinical trial'',\newline ``urinary retention clinical trial'',\newline ``Foley catheter clinical trial'',\newline ``high-dose steroids clinical trial'', \newline``hypertension clinical trial'',\newline ``chronic pain clinical trial'',\newline ``T-L spine clinical trial'',\newline ``unresectable tumor clinical trial'',\newline ``spinal cord conus mass clinical trial''
\end{tabular}
\caption{A few-shot prompting approach aiming at query generation by leveraging human-generated queries. System’s role:``You are a helpful medical assistant.''}
\label{tab:query_gen_refine_few_shot}
\end{table}
In terms of retrieval performance, this approach exhibits a slightly lower performance, i.e. Rprec equal to .168, compared to the QGMT approach, but the observed differences are not statistically significant.
Nonetheless, utilizing previously generated examples did not contribute to the generation of improved queries for unseen patient cases.
This outcome could be attributed to the choice of query examples or even the order in which these examples are included in the prompt, as suggested by previous studies~\cite{DBLP:conf/acl-deelio/LiuSZDCC22,lu-etal-2022-fantastically}.
Additionally, this approach increases the processing cost per patient case since it requires handling a bigger number of tokens in each system interaction.
In conclusion, this few-shot approach did not reach the expected retrieval performance compared to the majority of the zero-shot approaches presented in Table~\ref{tab:exp_ranking}.
\subsection{Observing the generation of possible hallucinated content}
\label{sec:hallu_observation}
During our experimentation, we utilized the zero-shot prompting approach presented in Table~\ref{tab:clinical_topics_generation}.
In this approach, ChatGPT is instructed to generate a list of potential clinical trials in which a selected patient may be eligible to participate.
The model has the flexibility to utilize its own knowledge and the information provided in the medical note to generate these recommendations.
In addition, the model is assigned a domain-specific role and is provided with a description of the task. 
\begin{table}[!htbp]
\setlength\tabcolsep{0.5em}
\begin{tabular}{p{8cm}|p{8cm}}
\multicolumn{1}{c|}{\centering Prompt Text} &
\multicolumn{1}{c}{\centering Model's output (example)} 
\\ \hline
You have access to a clinical medical note of a patient. 
Make a list of all possible clinical trials that this patient may be eligible for based on the information mentioned in the clinical medical note.
Consider all of the patient's information and exploit your own knowledge.
Report only the list of clinical trial topics. 
Add to each topic their associated MeSH terms.
Do not explain or elaborate and do not mention the term 'clinical' or 'trial'.
\newline\newline
Clinical note: {\textit{Clinical Note}}
&
Immunotherapy for anaplastic astrocytoma - MeSH terms: Immunotherapy, Anaplastic Astrocytoma\newline
Pain management in patients with anaplastic astrocytoma - MeSH terms: Pain Management, Anaplastic Astrocytoma\newline
Treatment options for unresectable anaplastic astrocytoma - MeSH terms: Treatment, Unresectable Anaplastic Astrocytoma\newline
Management of complications of anaplastic astrocytoma, such as urinary retention - MeSH terms: Complications, Anaplastic Astrocytoma, Urinary Retention\newline
Novel treatments for anaplastic astrocytoma - MeSH terms: Treatment, Novel, Anaplastic Astrocytoma\newline
Steroid-sparing treatment options for anaplastic astrocytoma - MeSH terms: Treatment, Steroid-Sparing, Anaplastic Astrocytoma\newline
Radiation therapy options for anaplastic astrocytoma - MeSH terms: Radiation Therapy, Anaplastic Astrocytoma\newline
Chemotherapy options for anaplastic astrocytoma - MeSH terms: Chemotherapy, Anaplastic Astrocytoma
\end{tabular}
\caption{A prompting approach aiming to generate a list of possible clinical trials in which a given patient can be eligible for. System’s role:``You are a helpful medical assistant.''}
\label{tab:clinical_topics_generation}
\end{table}
Regrading the observed retrieval performance in the TREC 2021 benchmark collection, this approach achieved an Rprec value equal to .167, which is still higher that other approaches presented in our study. However, the obtained P@10 (.224) and P@25 (.194) are lower compared to the other approaches.
These decreases in retrieval performance could be attributed to the model omitting important patient-related information from its responses, such as the absence of terms like "tumor" and "Avastin," among others.
Also, it is plausible that the generated topics may contain hallucinated content, i.e. topics that are not possible to exist or non-existing MeSH terms as reported in~\cite{DBLP:journals/corr/abs-2302-03495}. 
However, it is worth mentioning that due to the authors' lack of medical expertise, it is challenging to assess the accuracy of these generated topics, but we consider this research direction as potential future work.
On the other hand, it is possible that the model has identified new and previously unjudged clinical trials in which the considered patient could potentially participate.
Nevertheless, it is important to note that this approach grants the model considerable freedom in generating responses, which simultaneously renders the responses unpredictable and necessitates significant human supervision.
\section{Additional prompts for information extraction and expansion}
\label{sec:appendb}
As described in Section~\ref{sec:conclusions}, our findings indicate that when instructing ChatGPT to expand a list of keywords or refine generated queries using synonyms, it has a tendency to introduce terms with broader semantic meanings, resulting in topical drift and a subsequent decline in performance.
To address this, we conducted experiments with a prompting approach that involved providing ChatGPT with a domain-specific role, a comprehensive task description, and explicit instructions for keyword expansion (within a single system interaction).
Specifically, the model was directed to first extract all keywords describing the patient's information and then expand each keyword by incorporating two related medical concepts or synonyms.
The prompt that implements the aforementioned process is presented in Table~\ref{tab:explicit_query_expansion}.
\begin{table}[!htbp]
\setlength\tabcolsep{0.5em}
\begin{tabular}{p{7cm}|p{8cm}}
\multicolumn{1}{c|}{\centering Prompt Text} &
\multicolumn{1}{c}{\centering Model's output (example)}  \\ \hline
Given a medical clinical note your task is to extract all possible keywords that describe the patient's medical information without elaborating.
Then, for each keyword, provide two related medical concepts or synonyms based on your knowledge of the medical domain, without elaborating.
Your answer should consist of a comma-separated list of maximum 30 keywords that exhaustively describe the patient's medical information, with each keyword followed by its two related medical concepts or synonyms.
The keywords should not include duplicates and should not be elaborated on beyond the two related concepts or synonyms.
\newline\newline
Clinical note: {\textit{Clinical Note}}
&
anaplastic astrocytoma, brain tumor, spinal tumor
\newline spine, spinal cord, central nervous system
\newline lower extremity weakness, leg weakness, paralysis
\newline urinary retention, bladder dysfunction, incontinence
\newline Foley catheter, indwelling catheter, urinary catheter
\newline high-dose steroids, corticosteroids, glucocorticoids
\newline hypertension, high blood pressure, hypertensive
\newline chronic pain, persistent pain, long-term pain
\newline T-L spine, thoracolumbar spine, thoracic spine
\newline unresectable, inoperable, non-surgical
\newline radiation, radiotherapy, irradiation
\newline progressive, worsening, advancing
\newline right knee weakness, knee dysfunction, knee impairment
\newline difficulty walking, gait disturbance, ambulation difficulty
\newline right anterior thigh numbness, thigh sensory loss, leg paresthesia
\newline MRI, magnetic resonance imaging, imaging scan
\newline spinal cord conus mass, spinal cord tumor, tumor in conus medullaris
\newline biopsied, tissue sample, histopathology
\newline temozolomide, chemotherapy drug, alkylating agent
\newline field radiation, localized radiation, targeted radiation
\newline t10-l1, thoracic-lumbar region, specific spinal segments
\newline 11 cycles, treatment rounds, therapy sessions
\newline CPT-11, irinotecan, topoisomerase inhibitor
\newline Avastin, bevacizumab, anti-angiogenic agent
\end{tabular}
\caption{A prompting approach aiming to explicitly guide ChatGPT to perform query expansion for particular terms.
System’s role:``You are a helpful medical assistant.''}
\label{tab:explicit_query_expansion}
\end{table}
This approach does not demonstrate improved performance when compared to the BM25 baseline in the TREC 2021 benchmark collection.
It achieves an Rprec value of .151, indicating its lack of effectiveness in improving the baseline performance.
However, it is noteworthy that this approach showcases ChatGPT's capability to adhere to explicit instructions pertaining to query expansion.
Given this finding, we believe that future studies should explore in this direction, as it presents an opportunity for conducting fully interpretable and targeted query expansion.
Finally, by exploring and refining this approach, researchers can potentially enhance the effectiveness and interpretability of query expansion techniques.
\end{document}